\documentclass[fleqn,10pt]{wlscirep}

\usepackage[utf8]{inputenc}
\usepackage[T1]{fontenc}
\usepackage{bm}

\usepackage[style=numeric-comp,backend=biber,sorting=none]{biblatex}
\usepackage{xcolor,pifont} 
\usepackage{multirow} 
\usepackage{amsmath} 
\usepackage{graphicx}
\usepackage{wrapfig} 
\graphicspath{{images/}}
\usepackage{float}
\usepackage{comment}
\usepackage{blindtext}
\usepackage{array}    
\usepackage{subfiles}
\usepackage[british,UKenglish]{babel}
 \usepackage{csquotes} 

\DeclareFieldFormat{labelnumberwidth}{\mkbibbrackets{#1}}

\usepackage{verbatim}

\title{From FLOPs to Footprints: The Resource Cost of Artificial Intelligence}
\author[1*]{Sophia Falk}
\author[2]{Nicholas Kluge Corrêa}
\author[3]{Sasha Luccioni}
\author[4]{Lisa Biber-Freudenberger}
\author[1]{Aimee van Wynsberghe}

\affil[1]{Sustainable AI Lab, Institute for Science and Ethics, Bonn University, Germany}
\affil[*]{corresponding author: falk@iwe.uni-bonn.de}
\affil[2]{Center for Science and Thought, Bonn University, Germany}
\affil[3]{Hugging Face}
\affil[4]{Center for Development Research, Bonn University, Germany}

\begin{document}
\begin{abstract}

As computational demands continue to rise, assessing the environmental footprint of artificial intelligence (AI) requires moving beyond energy and water consumption to include the material demands of specialized hardware. This study quantifies the material footprint of AI training by linking computational workloads to physical hardware needs. The elemental composition of the Nvidia A100 SXM 40 GB graphics processing unit (GPU) was analyzed using inductively coupled plasma optical emission spectroscopy (ICP-OES), which identified 32 elements. The results show that AI hardware consists of approximately 90\% heavy metals and only trace amounts of precious metals. The elements copper, iron, tin, silicon, and nickel dominate the GPU composition by mass.
In a multi-step methodology, we integrate these measurements with computational throughput per GPU across varying lifespans, accounting for the computational requirements of training specific AI models at different training efficiency regimes. Scenario-based analyses reveal that, depending on Model FLOPs Utilization (MFU) and hardware lifespan, training GPT-4 requires between 1,174 and 8,800 A100 GPUs, corresponding to the extraction and eventual disposal of up to 7 tons of toxic elements. Combined software and hardware optimization strategies can substantially reduce material demands: increasing MFU from 20\% to 60\% lowers GPU requirements by $\approx$67\%, while extending lifespan from one to three years yields comparable savings; implementing both measures together reduces GPU needs by up to 93\%. Our findings highlight that incremental performance gains, such as those observed between GPT-3.5 and GPT-4, come at disproportionately high material costs. The study underscores the necessity of incorporating material resource considerations into discussions of AI scalability, emphasizing that future progress in AI must align with principles of resource efficiency and environmental responsibility.
\\
\end{abstract}

\maketitle
\section{Introduction}

Newspaper headlines such as `The world is running out of resources for IT' \cite{nasir2025world} and `Global shortage in computer chips reaches crisis point' \cite{sweney2021global} reflect a growing concern regarding the material constraints of the Fourth Industrial Revolution. At the center of this crisis is the semiconductor shortage, rooted in disruptions within the silicon supply chain. The previously significant semiconductor shortage occurred in 2021 during the COVID pandemic \cite{sweney2021global, kharpal2024surging}, and now a surge in demand for hardware accelerators for deep learning workloads could lead to the next global chip shortage \cite{kharpal2024surging}. It is important to note that the infrastructure powering emerging technologies, including artificial intelligence (AI)\footnote{By `AI', here we refer specifically to large-scale deep learning systems, such as those used for generative AI applications like chatbots.}, is built on far more than just silicon. Data centers, which form the physical backbone of emerging technologies, rely on a wide array of increasingly scarce and valuable materials, including rare earth elements, tantalum, and cobalt \cite{unctad2020digital}. While there is an increasing awareness of the environmental implications and supply chain constraints tied to the extraction of critical resources for `green' technologies and the energy sector, e.g. lithium for electric vehicles \cite{unep2019gobbling} or cobalt for renewable energy storage systems \cite{unep2024bend, iea2025executive}, there is a lack of equivalent scrutiny when it comes to the material demands of AI. As AI becomes a defining force of the ongoing Fourth Industrial Revolution, it is essential to identify and understand the `AI materials' that constitute the infrastructure and materiality of AI.\\
This becomes more urgent as global investments in digital infrastructure continue to accelerate. The rapid expansion of AI capabilities has driven unprecedented demand for high-performance computing, prompting substantial capital flows into data center construction and modernization. In 2024 alone, Alphabet, Microsoft, Meta, and Amazon collectively invested approximately \$246 USD billion in AI and data center development \cite{uddin2025big}. Projections indicate that this trajectory will continue: according to McKinsey, AI-ready data center capacity is expected to grow at an average annual rate of 33\% between 2023 and 2030, with AI workloads accounting for nearly 70\% of total data center demand by the end of the decade \cite{mckinsey2025ai}.
Despite this rapid expansion and increasing electricity demand, data centers accounted for only about 1.5\% of global electricity consumption in 2024 \cite{iea2024energy}, representing a comparatively modest share within the energy sector. Although the accelerating energy use associated with AI development and use is not negligible, other major drivers, such as electric vehicles, air conditioning, and electricity-intensive manufacturing, at present have a greater impact on overall global electricity consumption \cite{spencer2025what, iea2024world}. 
Data centers' current relatively contained global share is largely attributable to continuous improvements in energy efficiency, including advancements in cooling, workload management, and server performance. Nevertheless, the future trajectory of AI's energy demand, driven by its rapid development, ongoing trend of expanding model sizes, increasing adoption, and potential rebound effects remains uncertain and necessitate careful attention.\\
Efficiency improvements are closely intertwined with increased accessibility. Advances in model optimization have not only made AI systems more computationally efficient but also more widely deployable across diverse hardware environments, lowering the barrier to entry for AI use. Specific techniques, such as knowledge distillation\footnote{The compression of large AI models into smaller versions that maintain high performance while significantly reducing computational requirements \cite{medium2025small}.} and quantization\footnote{The reduction of parameter and computation precision (e.g., from 32-bit floating point to 8-bit integers), decreasing memory usage and increasing inference speed while maintaining accuracy \cite{lang2024comprehensive}.}, reduce model size and resource demands, improving both efficiency and deployability \cite{medium2025small}. However, these gains in efficiency and accessibility have simultaneously fueled broader AI adoption, ultimately reinforcing the demand for more infrastructure and, by extension, more hardware and material inputs \cite{luccioni2025REBOUND}. This is not a new phenomenon. As Jevon's Paradox, first articulated in 1866, demonstrated, increased efficiency often paradoxically leads to greater consumption instead of resource savings \cite{jevons2018coal}.\\
The assessment of AI's growing resource demands primarily focuses on its energy and water consumption. Energy analyses highlight the intensive computational requirements of training and deploying large-scale models \cite{deVries2023growing, luccioni2024powerHungry}, along with the substantial electricity demand of data center cooling systems\cite{zhang2021DCcooling}. These findings push current debates on net-zero pathways for AI, energy provision challenges, and decarbonization of electricity grids \cite{Xiao2025aiNetZeroPath}. Consequently, factors such as the geographic distribution of data centers, global trajectories of AI-related energy consumption, and energy infrastructure constraints play a central role in shaping the long-term carbon footprint of AI operations \cite{chen2025AIenergy}. More recently, water consumption has also gained attention as a critical dimension of AI's environmental footprint \cite{li2025makingAIlessthirty}. 
However, the two dimensions of carbon and water alone do not capture the full scope of AI's environmental footprint. As the demand for computational power continues to rise, a more comprehensive understanding of the physical footprint must also encompass AI's specialized hardware, hence including the materials that enable its operation.\\
The Nvidia A100 Graphics Processing Unit (GPU), for instance, is known to have played a transformative role in AI development \cite{leswing2023meet}. Following its release in 2020, the A100 quickly became a cornerstone of data center infrastructures at companies like OpenAI, Google, Microsoft, and Meta \cite{leswing2023meet, heath2024mark}. This was due to the A100's substantially improved performance, which has revolutionized AI model training and deployment.
The performance of a GPU can be measured by the number of floating-point operations it can perform per second (FLOPs). A FLOP is a single mathematical calculation involving decimal numbers, such as addition or multiplication. FLOPs per second measure how many of these calculations a processor can perform, making it a standard metric for describing the speed and capacity of GPUs, as well as for estimating the computational cost of training AI models. Capable of delivering up to 312 teraFLOPs (in BF16 precision) for deep learning workloads, the A100 provided a significant performance increase over its predecessor, the Volta series, with an approximately threefold increase in FP16 and sixfold in FP32 \cite{nvidia2020ampere}.\\
Beyond measuring GPU performance, FLOPs have emerged in regulatory contexts. The EU AI Act, for example, classifies high-risk AI models based on training compute budget, specifically defining general-purpose AI (GPAI) models as presenting systemic risks if their cumulative training exceeds 10$^{25}$ FLOPs \cite{eu2024regulation}. However, FLOPs go beyond computational intensity; they can also serve as a proxy for the material footprint of AI training. More FLOPs typically require more parameters, data, and training iterations, and consequently, greater hardware utilization.\\ 
To bridge the gap between digital computation metrics and physical resource consumption, we use the results from Falk et al. \cite{falk2025morethan} life cycle assessment's (LCA) background data, which presents the elemental composition of AI hardware using inductively coupled plasma optical emission spectroscopy (ICP-OES) on the Nvidia A100 SXM 40GB GPU to qualify and quantify the elements embedded in this critical infrastructure component. Second, we develop a framework linking computational demands (measured in FLOPs) to material consumption. This is achieved by mapping model training requirements to the Nvidia A100's operational capacity over its lifespan.\\
Thus, this study aims to establish a direct connection between the resource requirements of AI models and model performance metrics.
Our findings provide empirical evidence to inform technology policy and sustainable AI development toward a resource-conscious direction. By focusing on GPUs, a critical component at the core of any AI infrastructure, we reveal new insights into the underexplored link between the computational demands of AI models and the material extraction and waste generation they entail at the hardware level.

\section{Method}

The methodological framework of this study combines imported empirical data from prior physical component analysis with a newly developed approach for translating computational demand (in FLOPs) into estimates of material resource consumption. Specifically, we draw on results from our previous work, which involved the systematic disassembly and documentation of an Nvidia A100 SXM 40GB GPU, followed by ICP-OES analysis to quantify its elemental composition. Building on that established physical characterization, the present study introduces a method for estimating the maximum computational output a single Nvidia A100 GPU can deliver over different operational lifetimes. These computational capacity estimates are compared with the FLOP requirements of training selected AI models under various training-efficiency assumptions. From this comparison, we derive multiple scenarios for the number of GPUs needed to train a given model. Finally, these GPU-scaling scenarios are combined with the imported material composition dataset, enabling a scenario-based assessment of the material demands associated with training large-scale AI systems.

\subsection{Elemental Analysis Data}

\begin{wrapfigure}{r}{0.5\textwidth}
    \centering
    \includegraphics[width=0.5\textwidth]{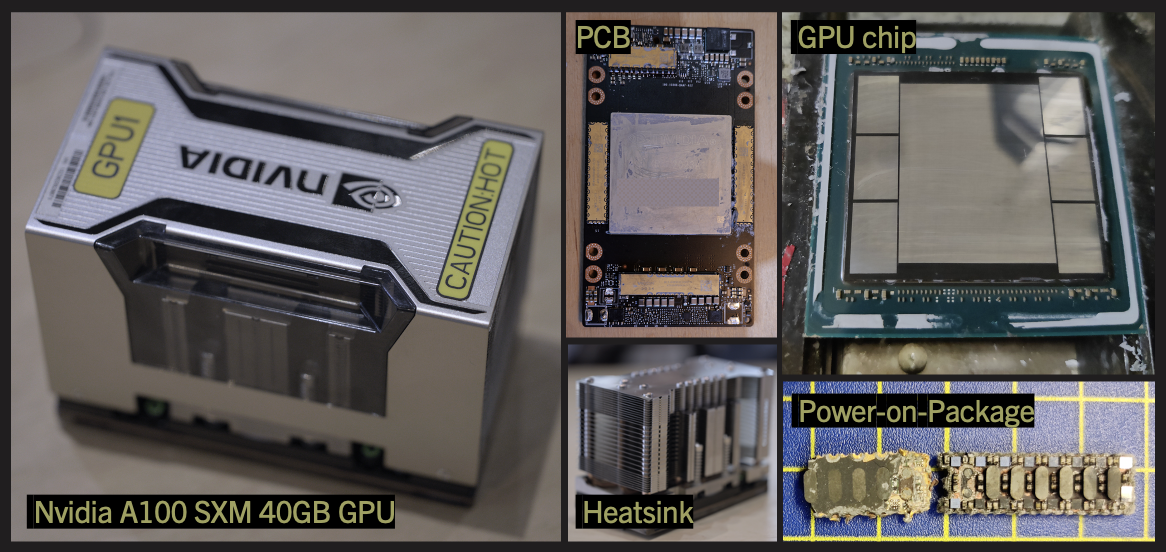}
    \caption{Annotation of the Nvidia A100 SXM 40 GB (left picture). For the chemical analysis conducted via ICP-OES, the GPU was disassembled and categorized into four main component categories: the printed circuit board (PCB) (top center), the heatsink (bottom center), the main GPU chip (top right), and the current regulators Power-on-Package (PoP) assemblies (bottom right) (author pictures).}
    \label{fig:annotated_components}
\end{wrapfigure}

The Nvidia A100 SXM 40GB GPU is disassembled into five component groups - casing, heatsink, printed circuit board (PCB), Power-on-Package (PoP), and GPU chip (VRAM and GPU die) (see Figure \ref{fig:annotated_components}) - and subsequently subjected to elemental characterization via ICP-OES. While the elemental composition dataset originates from the same empirical analysis reported in Falk et al. (2025) \cite{falk2025morethan}, that earlier work focused primarily on integrating these data into a LCA. In this section, we shift the emphasis from the LCA lens to the analytical procedures and results of the elemental analysis, providing methodological detail and interpretation not included in the prior publication.\\
\noindent
First, the sample preparation involves manual grinding to achieve particle sizes below 10 cm, followed by pyrolysis at 500°C for 2 hours to remove plastic components. The resulting residue is sequentially ground to obtain final particle sizes of 20-200 $\mu$m. Random quartering between grinding steps reduces sample quantities to 1.2 g per mineralization, with samples dried at 100°C for 24 hours before analysis.\\
Three distinct mineralization procedures are employed using an ANTON PAAR multiwave pro mineralizer to ensure comprehensive metal detection. The first procedure utilizes an oxidizing medium (H$_2$O$_2$ + HNO$_3$) optimized for silver quantification. The second employs a two-step process combining aqua regia with hydrogen peroxide, followed by nitric acid treatment. The third procedure used tetrafluoroboric acid and nitric acid to detect difficult-to-analyze elements, including silicon, tantalum, and gallium. All mineralizations are conducted at 240°C and 60 bar pressure over 1 hour (10 minutes temperature rise, 50 minutes plateau).\\
Post-mineralization solutions are centrifuged at 4000 rpm for 7 minutes and analyzed using Agilent ICP-OES 5100. Analysis employed both qualitative IntelliQuant screening (detection error <10\%) and quantitative methods using calibration curves with a minimum of three reference wavelengths per element to ensure accurate elemental composition determination.

\subsection{GPU Throughput and AI model Demands}

To link GPU throughput to the demands of training a given AI model, we assume a sequential training model in which the AI model is trained exclusively on a single GPU until the device reaches the end of its operational life, at which point the training is transferred to the following GPU. While this sequential training approach does not reflect real-world parallel training practices due to time and efficiency limitations, it provides a straightforward framework for estimating the total number of GPU hardware requirements based on the cumulative compute necessary to train an AI model.\\
The core calculation compares the GPU's throughput (i.e., the deliverable computational capacity) over its lifespan with the training compute demand of a given AI model. Two key variables highly influence the outcome of this comparison: (1) the assumed operational lifespan of the GPU and (2) the model training efficiency, defined as the proportion of delivered compute that effectively contributes to model training. Both factors are examined in detail in the following sections.

\subsubsection{Nvidia A100: Computational Throughput}

Technical documentation and industry analyses provide insights into the lifespan of data center GPUs. The physical durability of GPUs, such as the A100, is affected by component wear, workload intensity, maintenance, and cooling design \cite{grattafiori2024llama}. 
Anecdotal reports suggest that GPUs may last up to 5 years under heavy use and up to 7 years under moderate use \cite{ohiri2025nvidia, cybersided2025how}. However, empirical studies and technical reports indicate that the typical operational lifespan of GPUs in industry data centers is relatively short, around 3 years \cite{grattafiori2024llama,ostrouchov2020gpu}. In 2024, a Meta report on training the LLaMa 3 (405B) model found that GPU-related hardware issues accounted for a significant portion of unexpected training interruptions. Specifically, 58.7\% of all such incidents were linked to GPU failure, with faulty units alone accounting for 30.1\% of disruptions. Other common issues included memory or thermal failure \cite{grattafiori2024llama}. The report characterizes GPU failure as any event, hardware - or memory -related, that disrupts normal operations and requires human or automated intervention, such as GPU repair or replacement, to resume training \cite{grattafiori2024llama}. These frequent hardware failures imply that individual components have a relatively short mean time between failures in intensive training environments \cite{grattafiori2024llama}. Continuous, high-utilization workloads, especially in large-scale parallel training, accelerate GPU degradation. Supporting this, Ostrouchov et al. \cite{ostrouchov2020gpu} conducted a long-term study on a cluster of 18,688 Nvidia GPUs over nearly seven years. They found that the average time to hardware failure requiring intervention, such as replacement, is clustered around 2.8 years per GPU, reinforcing the relatively short lifespan of GPUs in demanding environments \cite{ostrouchov2020gpu}. An often-cited statement by an anonymous AI architect at Google echoes this view, stating that the maximum lifespan of data center GPUs at current utilization levels is 3 years. More specifically, given continuous high utilization rates, the expected lifespan of a GPU typically falls between 1 and 2 years, rarely exceeding 3 years \cite{techfund2024ai, shilov2025datacenter}. While improved maintenance and lower workload intensity can extend GPU lifespans, doing so may introduce additional costs and operational constraints that cloud providers are often unwilling to bear, since it effectively limits the amount of compute they can offer over time. Consequently, high GPU utilization remains the norm to maximize performance \cite{ostrouchov2020gpu} and return on investment \cite{shilov2025datacenter}.
Therefore, for our estimations, we provide results for lifespans of 1, 2, and 3 years, which we believe are realistic for this type of hardware.\\
Building on these lifespan assumptions, we next quantify the GPU's computational capacity over time. By assessing the A100 GPU's FLOPs performance using bfloat16 (BF16) precision and projecting this across different lifespan scenarios, we can estimate the total computational throughput the GPU is capable of delivering over its lifetime (see Table \ref{tab:theoretical throughput_annual}).
The peak BF16 throughput of the A100 (both 40 GB and 80GB models) is 312 teraFLOPs per second\footnote{Higher FLOPs are achieved only for sparsity, which is usually not applicable for most use cases involving deep learning workloads.} \cite{ohiri2025nvidia, bahdanau2025flops}. Based on continuous annual operation, the total yearly computational throughput can be calculated as:

\begin{equation}
\text{Annual Computational Throughput} = (312 \times 10^{12}) \times (365 \times 24 \times 60 \times 60 ) = 9.8 \times 10^{21} \text{ FLOPs}
\label{eq:annual_flops}
\end{equation}
Following Equation \ref{eq:annual_flops}; at 100\% utilization over one year, this corresponds to 9.8 x 10$^{21}$ FLOPs annually.\\ 
\begin{table} [H]
    \centering
        \caption{Total amount of maximum theoretical throughput expressed in FLOPs processed per A100 (BF16) over its operational lifespan, assuming three lifespan scenarios.}
    \begin{tabular}{cc}\toprule
         Hardware lifespan (years)& Theoretical Computational Throughput\\\midrule
         1& 9.8 x 10$^{21}$\\
         2& 1.9 x 10$^{22}$\\
         3& 2.9x 10$^{22}$\\ \bottomrule
    \end{tabular}
    \label{tab:theoretical throughput_annual}
\end{table}

\subsubsection{AI Model Training: Computational Requirements}

For this study, we examine the computational requirements of models trained using Nvidia A100 GPUs specifically (see Table \ref{tab:mAImodel_overview}). Since not all models report computational requirements in terms of FLOPs, we estimate them from available model specifications. For dense transformer architectures, a widely used heuristic, proposed by Kaplan et al. (2020), assumes approximately 6 FLOPs per parameter per token during training \cite{kaplan2020scaling}. The corresponding computational budget is expressed as:
\begin{equation}
    \text{Compute Budget} = 6 \times N \times D
    \label{eq:compute_budget}
\end{equation}
\noindent
Where $N$ is the number of model parameters and $D$ the number of training tokens. All AI models considered in this paper, except GPT-4, are or strongly presumed to be dense transformer models.\\
For sparse architectures, such as Mixture-of-Experts (MoE) models, only a subset of parameters is activated per forward pass\cite{fedus2021switch}. To estimate the compute budget for MoEs, Equation \ref{eq:compute_budget} remains valid; however, only the active parameters are considered for $N$.\\
The architecture of GPT-4 remains officially undisclosed. OpenAI's GPT-4 technical report states that the model is Transformer-based but provides no further details on its size, number of parameters, or architectural configuration \cite{openai2024gpt4technicalreport}. 
Nevertheless, multiple technical analyses strongly suggest that GPT-4 employs a MoE architecture rather than a dense Transformer design. 
This hypothesis is supported by two observations: First, inference latency is significantly lower than would be expected for a dense model of comparable capacity\cite{patelWong2023GPT-4Architecture, Erdil2024MoEinference}. And second, the computational and financial costs of training and inference would be prohibitively high if GPT-4 were fully dense \cite{patel2023ScalingWall}. 
Several reports propose different MoE configurations for GPT-4. One such report suggests that GPT-4 consists of eight experts, each with approximately 220 billion parameters, with two experts active per forward pass, totaling about 1.76 trillion parameters \cite{Fischer2025MoEExplained}. Another analysis proposes 16 experts, each containing 111 billion parameters, with two experts active \cite{SemiAnalysis2023GPT-4Archi}. To account for uncertainty, we evaluate GPT-4 under multiple activation scenarios, similar to the MoE architectures of Ling- 1T, DeepSeek-V3, and Llama 4 Maverick. This scenario-based approach enables the estimation of GPT-4's computational and material footprints across a plausible range of active-parameter configurations.\\
These architectural assumptions serve as the foundation for estimating the hardware requirements associated with model training. Building on the computed training demands for both dense and sparse architectures, we next relate each model's total computational budget to the operational throughput and lifespan of individual GPUs to determine the number of units effectively required for training:\\
\begin{equation}
\text{Required GPUs} = \frac{\text{Compute Budget}}{\text{Annual Computational Throughput} \times \text{Lifespan}}
\label{eq:gpu_requirements}
\end{equation}
\noindent
Where the lifespan represents the expected productive duration in high-utilization data center environments.\\
These estimates assume continuous 24/7 operation at peak theoretical performance, i.e., 100\% model FLOPs utilization (MFU), which represents an idealized upper bound. In practice, more complex training setups (e.g., distributed multi-dimensional parallelism across multiple hosts), introduce latency, communication overheads, and input-output (I/O) bottlenecks that reduce effective utilization \cite{jiang2024megascale, zhao2024efficiently, jamil2025emlio, go2025characterizing}.

\subsubsection{Efficiency Adjustments for Real-World Training Scenarios}

Empirical studies indicate that real-world MFU values typically range from 20\% to 60\% of peak throughput \cite{chowdhery2022palm, pytorchLargeScale}. For instance, Meta reported 38 - 41\% during the training of LLaMa 3 (405B) \cite{fryepaid}, and 53\% for LLaMa 2 \cite{pytorchHighPerfLlama2}. At the same time, OpenAI achieved 19.6\% MFU for pre-training GPT-3 and approximately 34\% MFU for GPT-4, and Google reported 46.2\% for PaLM \cite{chowdhery2022palm}. MFU values above 50\% are generally regarded as indicative of highly optimized training \cite{jiang2024megascale}. 	
However, MFU is not consistently disclosed across models and can vary considerably depending on data center infrastructure, interconnect topology, networking bandwidth, storage I/O systems, software stack configurations, and operational practices \cite{jiang2024megascale, pytorchMaximizing}. To account for this variability and the lack of standardized MFU reporting, we present GPU requirements across a representative MFU range of 20\% to 50\% for the three hardware lifespan scenarios (1, 2, and 3 years). \\
Note that a lower MFU does not imply proportionally lower power consumption. Technical reports show that GPUs often operate at or near maximum power draw even at 20\%-50\% MFU \cite{behindthekeyboardLlama, chowdhery2022palm}. Since power draw and associated thermal dissipation are key drivers of component degradation and catastrophic failure \cite{tang2024brief, Ycui_2021_flexible}, reduced MFU does not extend GPU lifespan.\\
To account for the difference between theoretical and real-world GPU usage, we apply a scaling factor to compensate for training inefficiencies:

\begin{equation}
\text{Scaling factor} = \frac{1}{\text{MFU}} \quad \text{where MFU} \in [0.20, 0.60]
\label{eq:scaling_factor}
\end{equation}
\noindent
The adjusted GPU count is then computed as:\\

\begin{equation}
\text{GPU}_{\text{adjusted}} = \text{GPU}_{\text{required}} \times \text{Scaling factor}
\label{eq:gpu_adjusted_lifespan}
\end{equation}
\noindent
\\
These scenarios provide a comprehensive assessment of potential hardware requirements under varying operational conditions. The estimates remain conservative, as they consider only training time, MFU, and lifespan, excluding additional delays or training interruptions that would further increase resource use.

\subsection{Material Footprint of AI Model Training}

In the final step of the analysis, the material composition of a single GPU, determined through ICP-OES, is scaled by the number of GPUs required to train each AI model. The total material demand for each element is expressed as:
\begin{equation}
M_{\text{element}} = m_{\text{element}} \times GPU_{\text{adjusted}} (n)
\label{eq:material_footprint}
\end{equation}
\noindent
Where $m_{\text{element}}$ is the mass of a specific element detected in the analyzed GPU, and $GPU_{\text{adjusted}}(n)$ corresponds to the adjusted GPU requirement accounting for real-world MFU and hardware lifespan. Multiplying these terms yields the total elemental mass required for training a given AI model.\\
This computation is repeated for all 32 detected elements across various training scenarios to capture the uncertainty range in hardware utilization and lifespan assumptions. For models with reported MFU data, material requirements are reported for the nearest MFU increment combined with the 1-3 year lifespan range. Models lacking publicly available MFU values are assessed across the entire parameter range to ensure a comprehensive evaluation of their potential resource intensity.

\section{Results}

This section presents the computational requirements for training state-of-the-art AI models on the Nvidia A100 SXM 40GB GPU. While AI training typically occurs in large-scale data center environments supporting diverse digital workloads, this analysis isolates the material requirements directly attributable to GPU-based AI training. 
It is important to note that the results here reflect only the GPU unit. Broader infrastructure components, such as networking, storage, and cooling systems, are excluded from this analysis. Including these hardware components would substantially increase the total material footprint, and the values reported here should therefore be interpreted as conservative lower-bound estimates of AI's overall material demand.

\subsection{GPU requirement per AI model}

Using Equation \ref{eq:compute_budget}, we derive the computational budgets of eight large-scale dense Transformer models developed between 2022 and 2024, trained on Nvidia A100 GPUs (see Table \ref{tab:mAImodel_overview}). For GPT-4, we separately estimate a range of plausible computational budget scenarios based on MoE architectural assumptions (see Table \ref{tab:GPT4-MoE}).
As previously explained, for our estimations, training is assumed to run sequentially, with each GPU used continuously until it reaches the end of its productive life, at which point the following GPU takes over. This lets us directly estimate how many GPUs are effectively used during training and, thus, the material demand involved.
While it is also possible to estimate the proportional wear and tear across all GPUs used in parallel, the sequential modeling approach more effectively demonstrates the scale and intensity of hardware turnover implicit in contemporary AI training workloads.

\begin{table} 
    \centering
       \caption{Overview of large-scale dense Transformer models published between 2022 and 2024 that were trained on Nvidia A100 SXM hardware. The estimated number of FLOPs per model is derived from reported or inferred parameter and token counts. Where official data on parameters or token counts is unavailable, italicized values indicate estimates based on industry conventions and credible leaked data sources.}
    \begin{tabular}{llll}\toprule
         \textbf{Model}&  \textbf{Parameter}&  \textbf{Token}& \textbf{FLOPs}\\
         &($N$) & ($D$) & Compute Budget\\ 
         \midrule
         Amazon Titan TG-1&  200 B&  4 trillion& 4.8 x 10$^{24}$\\
         Mistral Large 2&  123 B&  \textit{2 trillion}& 1.48 x 10$^{24}$\\
         LLama 2&  70 B&  2 trillion& 8.4 x 10$^{23}$\\
         DeepSeekLLM&  67 B&  2 trillion& 8.04 x 10$^{23}$\\
         Bloom&  176 B&  366 billion& 3.86 x 10$^{23}$\\
         GPT-3.5&  175 B&  \textit{300 billion}& 3.15 x 10$^{23}$\\
         Falcon&  40 B&  1 trillion& 2.4 x 10$^{23}$\\
         Phytia&  12 B&  300 billion& 2.16 x 10$^{23}$\\ \bottomrule
    \end{tabular}
    \label{tab:mAImodel_overview}
\end{table}
 
\begin{table}[h]
\centering
\caption{The estimated GPT-4 MoE configurations are based on publicly discussed architectural hypotheses and disclosed MoE architectures assuming a training dataset of 13 trillion tokens. The computational budget is estimated using parameter activation ratios informed by the MoE architectures of Ling-1T, DeepSeek-V3, and Llama 4 Maverick. For the remainder of the study, we use the scenario laid out by SemiAnalysis as a conservative approach.}
\label{tab:gpt4_scenarios}
\begin{tabular}{@{}lcccc@{}}
\toprule
\textbf{Scenario} & \textbf{Total} & \textbf{Active} & \textbf{Token} &\textbf{FLOPs}\\ 
 & \textbf{Parameter} ($N$) & \textbf{Parameter} ($N$$^*$)& ($D$) &Compute Budget\\ 
\midrule
George Hotz & 1.76T & 440B  & 13 trillion &3.43 x 10$^{25}$\\
\textbf{SemiAnalysis} & \textbf{1.76T }& \textbf{222B} &  \textbf{13 trillion} & \textbf{1.73 x 10$^{25}$}\\
\midrule
Ling-1T-like & 1.76T & 50B&  13 trillion & 3.90 x 10$^{24}$\\
DeepSeek-V3-like& 1.76T & 37B& 13 trillion & 2.89 x 10$^{24}$ \\
Llama 4 Maverick-like& 1.76T & 17B&  13 trillion & 1.33 x 10$^{24}$ \\
\bottomrule
\end{tabular}
\label{tab:GPT4-MoE}
\end{table}
\begin{table} [H]
    \centering
    \caption{Total GPU requirements for training selected AI models under varying hardware lifespan (1–3 years) and model FLOPs utilization (MFU) scenarios (20\% lower bound, 50\% upper bound). Values are rounded up to the nearest integer.}
    \begin{tabular}{c|c|cc|cc|cc}
    \toprule
         \multirow{2}{*}{\textbf{Model}} & \multirow{2}{*}{\textbf{FLOPs}} & \multicolumn{2}{c|}{\textbf{Lifespan = 1}} & \multicolumn{2}{c|}{\textbf{Lifespan = 2}} & \multicolumn{2}{c}{\textbf{Lifespan = 3}} \\
         \cmidrule(lr){3-4} \cmidrule(lr){5-6} \cmidrule(lr){7-8}
         & & \textbf{MFU 20\%} & \textbf{MFU 50\%} & \textbf{MFU 20\%} & \textbf{MFU 50\%} & \textbf{MFU 20\%} & \textbf{MFU 50\%} \\
         \midrule
         GPT-4 & 1.73 × 10$^{25}$& 8,800& 3,520& 4,400& 1,760& 2,934& 1,174\\
         Amazon Titan & 4.8 × 10$^{24}$ & 2,439 & 976 & 1,220& 488& 814& 326\\
         Mistral Large 2 & 1.48 × 10$^{24}$ & 751& 301 & 377& 151& 251& 151\\
         LLama 2 & 8.4 × 10$^{23}$ & 427 & 171 & 214& 85.38 & 143& 57\\
         DeepSeekLLM & 8.04 × 10$^{23}$ & 409 & 164& 205& 82 & 137& 55\\
         BLOOM & 3.86 × 10$^{23}$ & 196 & 78 & 99& 40 & 66& 27\\
         GPT-3.5 & 3.15 × 10$^{23}$ & 160 & 64 & 80& 32& 54& 22\\
         Falcon & 2.4 × 10$^{23}$ & 122 & 49 & 61 & 25& 41& 17\\
         Pythia & 2.16 × 10$^{23}$ & 11 & 4 & 6& 3& 4& 2\\
         \bottomrule
    \end{tabular}
    \label{tab:GPU_count}
\end{table}
\noindent 
Using Equation \ref{eq:gpu_adjusted_lifespan}, we estimate the number of Nvidia A100 GPUs required to train each AI model based on three life span scenarios and applying a lower bound MFU of 20\% and an upper bound estimate of 50\% MFU (see Table \ref{tab:GPU_count}). GPU requirements scale with model complexity, i.e., compute budget. Training GPT-4 (based on the MoE SemiAnalysis scenario) with an MFU of 20\% and a hardware lifespan of 1 year requires 8,800 A100 GPUs, decreasing to approximately 2,934 GPUs with a hardware lifespan of 3 years. Improving the MFU to 50\% reduces the range from 3,520 GPUs (1-year lifespan) to 1,174 GPUs (3-year lifespan). GPT-4 requires the highest number of GPUs, followed by Amazon Titan (2,439 to 326 GPUs), Mistral Large 2 (752 to 151 GPUs), LLaMa 2 (427 to 57 GPUs), DeepSeekLLM (409 to 55 GPUs), BLOOM (196 to 27 GPUs), GPT-3.5 (160 to 22 GPUs), and Falcon (122 to 17 GPUs). Pythia demonstrates the lowest requirements (11 to 2 GPUs), illustrating the substantial variation in computational demands across different model sizes.\\  

\subsection{Elemental Composition of the Nvidia A100 GPU}

\begin{figure}
    \centering
\includegraphics[width=1\linewidth]{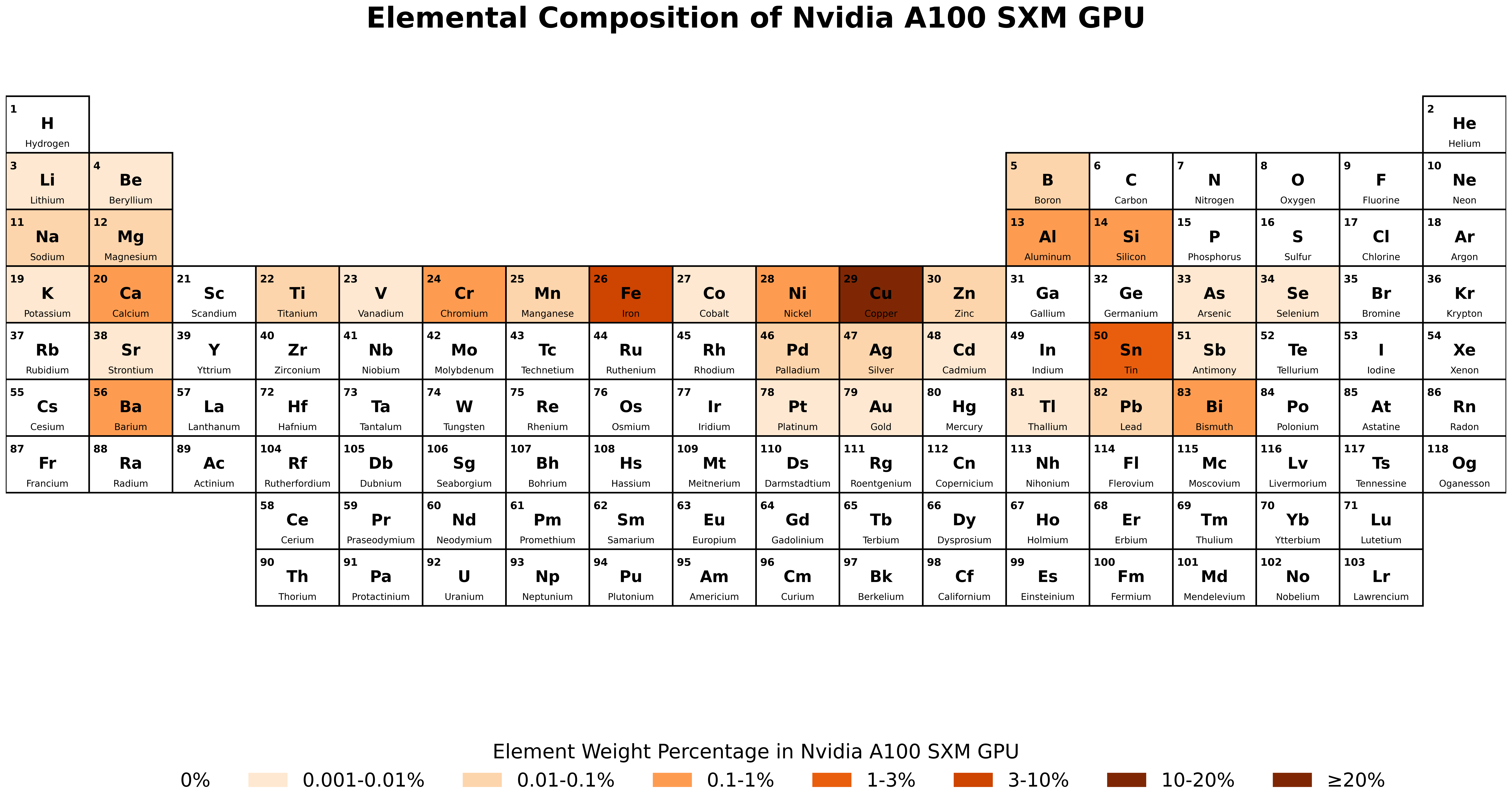}
    \caption{Proportion of elements in the Nvidia A100 SXM 40GB GPU (author illustration).}
    \label{fig:periodic_table}
\end{figure}

\begin{wrapfigure}{l}{0.5\textwidth}
    \centering
\includegraphics[width=0.52\textwidth]{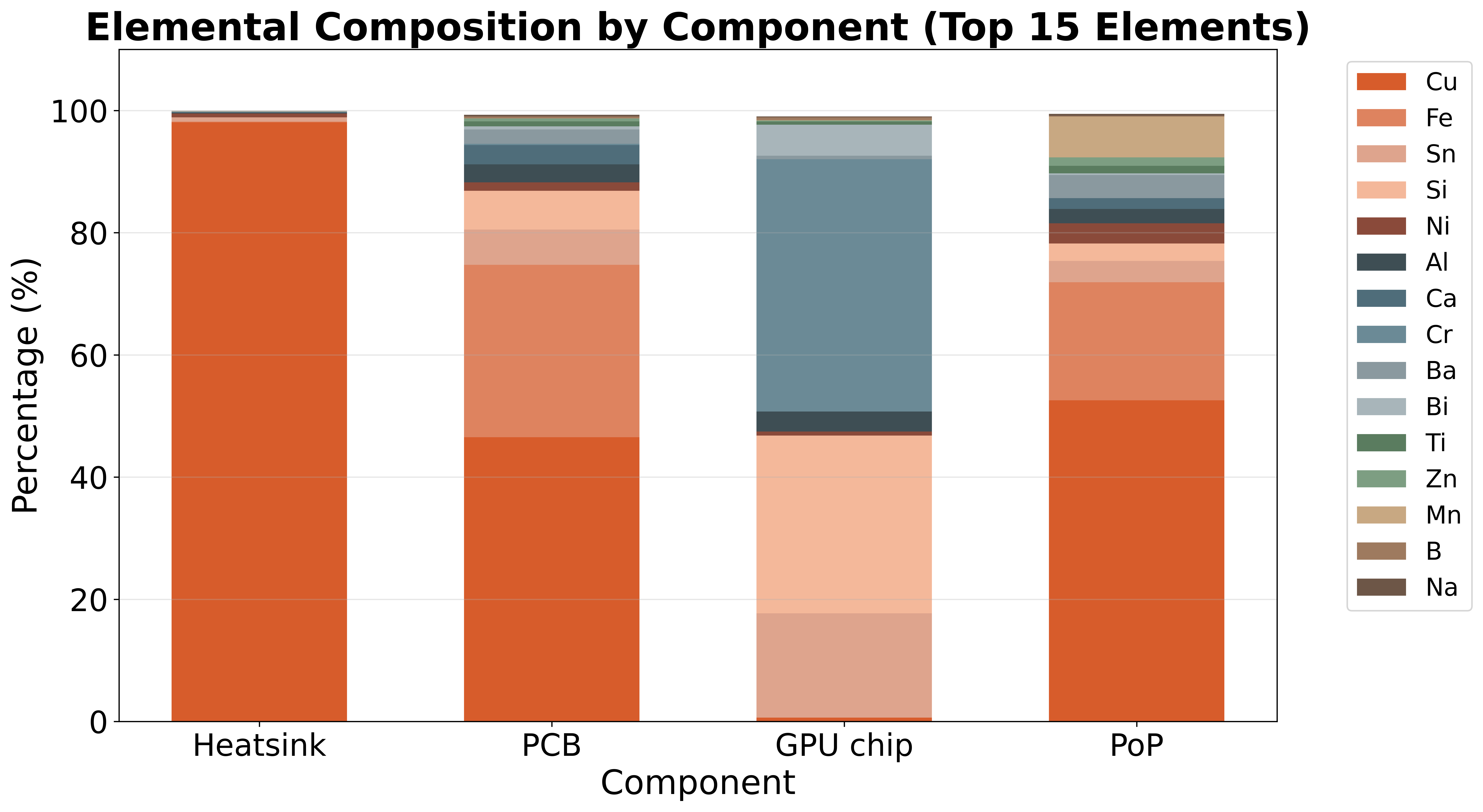}
    \caption{Elemental composition of the Nvidia A100 SXM GPU by component group: heatsink, PCB, main GPU (GPU chip + VRAM), and Power-on-Packages. Displayed are the top 15 elements by weight proportion (\% of total mass) (author illustration).}
    \label{fig:component_comp}
\end{wrapfigure}

The ICP-OES elemental analysis of the GPU identifies the presence of 32 elements from the periodic table (see Figure \ref{fig:periodic_table}). About 84\% of the present elements are metals, 12.5\% are metalloids, and one is a nonmetal. Four of the eight precious metals - gold, silver, platinum, and palladium - were detected, albeit primarily in trace amounts. Silver is the most abundant among them, with 0.55 grams per GPU. Consequently, the economic value of precious metals per individual GPU is relatively limited.\\
The quantities of elements vary widely, with copper being the most abundant at 1,374 grams, while beryllium registers as the least abundant at just 0.0000237829 grams (see Table \ref{tab:nvidia_a100}). The most dominant elements by mass are copper, iron, tin, silicon, and nickel. 
Beyond the quantitative elemental composition, the relative distribution of 32 elements across the four individual GPU components is examined.
The elemental composition differs considerably across the GPU components, specifically the heatsink, PCB, GPU chip, and Power-on-Packages (PoP) (see Figure \ref{fig:component_comp}).\\

\noindent
Figure \ref{fig:component_comp} shows the ranking of elements according to their total mass contribution across all components. The top 15 elements are visualized by their proportional contributions within each component. Copper and iron constitute substantial portions of the GPU's structural components, primarily in the heatsink and PCB. In contrast, silicon and nickel are predominantly concentrated in the functional components, namely the GPU chip and PoP.
The heatsink, the most substantial component by mass, is composed of 98.1\% copper, reflecting its primary role in thermal management. In contrast, the PCB exhibits a more heterogeneous elemental profile, comprising 46.5\% copper and 28\% iron, along with smaller proportions of silicon, tin, and calcium. The internal computing structure, the GPU chip, and PoP are characterized by higher elemental complexity. The PoP is composed of 52.6\% copper, 19\% iron, 6.6\% magnesium, with notable amounts of barium and zinc. The GPU chip primarily comprises 41\% chromium, 29\% silicon, 17\% tin, and smaller portions of bismuth and aluminum. 
The substantial amount of silicon within the GPU chip is primarily attributable to the large die size of the main GPU processor. Our analysis of the dismantled unit reveals that approximately 1,353 mm$^2$ of silicon area is integrated within a single 55 mm x 55 mm, 12-layer packed GPU inside the Nvidia A100 SXM. In comparison, its predecessor, the Nvidia Tesla V100, features a significantly smaller die area of 815 mm$^2$ \cite{nvidia2017volta}. To meet the computational demands of large-scale AI training, semiconductor devices are increasingly scaling their physical dimensions, either through larger die areas or advanced packaging that integrates multiple smaller dies. These trends lead to increased silicon consumption as AI models continue to grow. However, constraints such as the maximum reticle area, manufacturing costs, cooling challenges, and diminishing functional die yield impose practical limits on die scaling, a phenomenon often referred to as `area-wall' \cite{han2023big}. Among these limitations, thermal management emerges as a particularly critical challenge as chip size increases \cite{han2023big}, underscoring the need for enhanced cooling solutions and the consideration of waste heat utilization \cite{falk2025DC_wasteheat}. 
Hence, we suggest that meeting the computational demands of increasingly large AI models will necessitate scaling via additional GPU units rather than fewer, enhanced individual chip capabilities. This shift carries significant implications for material consumption. As AI models continue to grow, the full spectrum of resources required for manufacturing complete GPU units will be extracted, rather than primarily increasing silicon consumption by enlarging individual, more powerful chips.
\begin{table}[htbp]
\centering
\caption{Elemental composition of the Nvidia A100 SXM 40 GB GPU by component (heatsink, PCB, GPU chip, and PoP) and in total (mass in grams). Elements marked with (*) are either inherently toxic or present significant health hazards through exposure to their vapors and dusts during extraction and processing.}
\begin{tabular}{llrrrrr}
\hline
\textbf{Element} & \textbf{Abbr.} & \textbf{Heatsink} & \textbf{PCB} & \textbf{GPU chip} & \textbf{PoP} & \textbf{Total} \\
\hline
Silver & Ag & 4.32E-01 & 1.02E-01 & 1.13E-02 & 7.85E-03 & 0.553 \\
Aluminum & Al & 1.89E+00 & 4.33E+00 & 4.18E-01 & 2.65E-01 & 6.90E+00 \\
Arsenic (*)& As & 2.69E-02 & 5.92E-03 & 2.87E-04 & 3.75E-04 & 3.35E-02 \\
Gold & Au & 2.96E-02 & 1.01E-02 & 1.15E-03 & 1.71E-03 & 4.26E-02 \\
Boron & B & 2.77E-01 & 4.30E-01 & 6.04E-02 & 1.47E-02 & 7.82E-01 \\
Barium & Ba & 7.27E-02 & 3.47E+00 & 7.24E-02 & 4.31E-01 & 4.05E+00 \\
Beryllium (*) & Be & 1.35E-03 & 2.19E-04 & 2.22E-02 & 1.39E-05 & 2.38E-02 \\
Bismuth & Bi & 1.38E+00 & 7.43E-01 & 6.47E-01 & 3.18E-02 & 2.80E+00 \\
Calcium & Ca & 9.28E-01 & 4.65E+00 & 2.21E-05 & 2.02E-01 & 5.78E+00 \\
Cadmium (*) & Cd & 2.69E-03 & 1.10E-03 & 5.75E-04 & 6.95E-05 & 4.43E-03 \\
Cobalt (*) & Co & 4.04E-03 & 5.70E-03 & 9.72E-04 & 5.82E-03 & 1.65E-02 \\
Chromium (*) & Cr & 6.19E-02 & 3.43E-01 & 5.25E+00 & 1.33E-03 & 5.66E+00 \\
Copper (*) & Cu & 1.30E+03 & 6.87E+01 & 8.14E-02 & 5.99E+00 & 1.37E+03 \\
Iron & Fe & 1.58E+00 & 4.17E+01 & 1.75E-03 & 2.20E+00 & 4.55E+01 \\
Potassium & K & 2.69E-03 & 2.87E-02 & 8.84E-05 & 3.06E-04 & 3.18E-02 \\
Lithium & Li & 2.69E-03 & 4.39E-04 & 6.99E-02 & 5.56E-05 & 7.31E-02 \\
Magnesium & Mg & 9.83E-02 & 1.86E-01 & 9.22E-03 & 1.78E-02 & 3.11E-01 \\
Manganese & Mn & 8.62E-02 & 7.00E-02 & 4.40E-03 & 7.57E-01 & 9.18E-01 \\
Sodium & Na & 2.49E-01 & 3.88E-01 & 1.62E-02 & 3.73E-02 & 6.91E-01 \\
Nickel (*) & Ni & 8.55E+00 & 2.05E+00 & 8.17E-02 & 3.76E-01 & 1.11E+01 \\
Lead (*) & Pb & 1.27E-01 & 5.00E-01 & 4.71E-03 & 2.43E-02 & 6.56E-01 \\
Palladium & Pd & 4.85E-02 & 1.49E-01 & 9.50E-04 & 1.86E-03 & 2.00E-01 \\
Platinum & Pt & 4.98E-02 & 5.49E-03 & 8.40E-04 & 4.17E-04 & 5.65E-02 \\
Antimony (*) & Sb & 6.60E-02 & 1.73E-02 & 9.50E-04 & 7.78E-04 & 8.50E-02 \\
Selenium & Se & 6.06E-02 & 3.16E-02 & 5.30E-04 & 9.31E-04 & 9.37E-02 \\
Silicon & Si & 0.00E+00 & 9.37E+00 & 3.70E+00 & 3.27E-01 & 1.34E+01 \\
Tin & Sn & 9.20E+00 & 8.50E+00 & 2.17E+00 & 3.98E-01 & 2.03E+01 \\
Titanium & Ti & 8.08E-02 & 1.21E+00 & 6.42E-02 & 1.39E-01 & 1.49E+00 \\
Thallium (*) & Tl & 1.75E-02 & 4.17E-03 & 5.08E-04 & 6.12E-04 & 2.28E-02 \\
Vanadium & V & 8.08E-03 & 3.29E-03 & 2.65E-04 & 5.84E-04 & 1.22E-02 \\
Zinc (*) & Zn & 3.25E-01 & 6.69E-01 & 2.42E-02 & 1.60E-01 & 1.18E+00 \\
Strontium & Sr & 5.39E-03& --- & --- & --- & 5.39E-03\\
\hline
\end{tabular}
\label{tab:nvidia_a100}
\end{table}

\subsection{The Resource Cost of AI training}

\begin{figure}
    \centering
\includegraphics[width=1\linewidth]{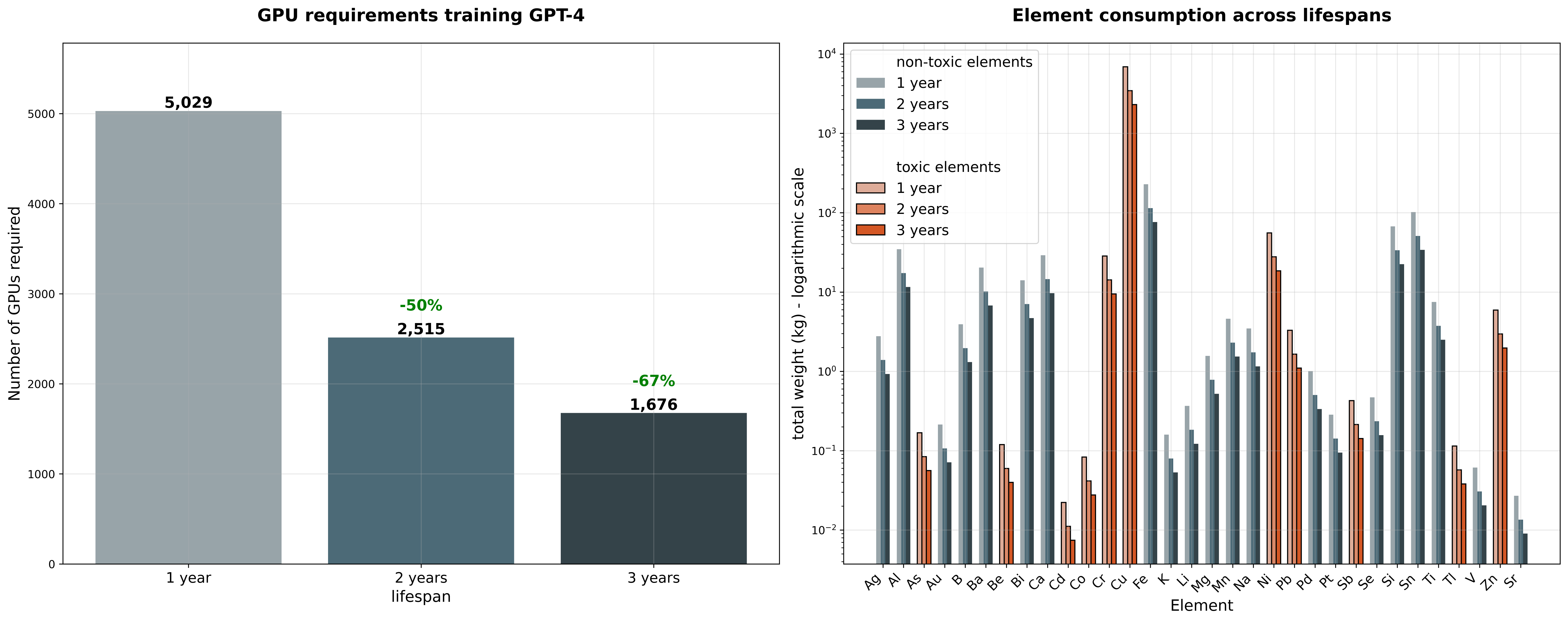}
    \caption{Estimated hardware and elemental requirement for training GPT-4 at 35\% MFU across varying hardware lifespan scenarios (1 to 3 years). Results are expressed in terms of the total number of GPUs required and the total elemental mass (kg) on a logarithmic scale. Extending the hardware's operational lifespan to 2 years halves the GPU demand to 2,515 GPUs, while a 3-year lifespan reduces requirements by approximately 67\% to 1,676 GPUs (author's illustration).}
    \label{fig:gpt-4_35MFU}
\end{figure}

Training GPT-4 with a reported MFU of around 35\% \cite{katerinaptrvGPT4Leaked} requires the computational capacity of approximately 5,029 A100 GPUs over a one-year hardware lifespan (see Figure \ref{fig:gpt-4_35MFU}). These computational demands translate directly into material impacts: one training round of this single model requires the extraction and eventual disposal of an estimated 7,003 kg of toxic materials.\\ 
\noindent
Electronic components contain numerous heavy metals, including arsenic (As), mercury (Hg), lead (Pb), cadmium (Cd), chromium (Cr), zinc (Zn), copper (Cu), nickel (Ni), antimony (Sb), cobalt (Co) and beryllium (Be), that pose acute ecological and health risks if released during mining, manufacturing or disposal at the end of their life \cite{wang2024analysis, wu2022antimonyBeCO, WHO2010toxic}. Many of these metals are classified as carcinogenic or highly toxic. Exposure through inhalation, dermal contact, or ingestion of contaminated water can lead to lung cancer, neurological impairment, gastrointestinal disorders, and other long-term health impacts \cite{wang2024analysis}. Toxic metals pose a health hazard in mining at varying concentrations depending on the metal; lead, for example, is dangerous at extremely low parts-per-billion (ppb) levels \cite{WHO2010toxic}. In developing countries, untreated or inadequately treated industrial wastewater and mining represent primary sources of metal pollution in freshwater systems\cite{dutta2021IndustrialWastewater}.\\
Our elemental analysis indicates that all of these toxic metals are present in the A100 GPU, accounting for around 93\% of the device's total mass. The dominant metal is copper (91\%), followed by nickel (0.74\%) and chromium (0.38\%). While the concentrations may appear modest at the level of a single unit, large-scale AI training requires thousands of GPUs, thereby magnifying the pressures of upstream extraction and the risks of downstream contamination across soil, air, and groundwater systems. In many African mining regions, concentrations of toxic metals in the soil and water substantially exceed WHO drinking water thresholds posing risks to communities \cite{ripanda202AfricaMining} (e.g. levels are above 10 $\mu$g/L for As and Pb, 3 $\mu$g/L for Cd, 20 $\mu$g/L for Cr and 2000 $\mu$g/L for Cu \cite{WHO2010toxic}.)\\
Assuming an MFU of 20\% and a 1-year GPU lifespan, training the nine models listed in Table \ref{tab:GPU_count} requires 13,315 GPUs. This corresponds to a material footprint of 19,940 kg extracted resources, with toxic materials comprising about 18,545 kg. Under more optimal conditions of a 3-year lifespan and 50\% MFU, material extraction decreases to about 2,740 kg, of which 2,550 kg remain toxic. However, these nine models represent only a fraction of the aggregate AI industry resource consumption.
To quantify sector-wide resource consumption and compare with other industries, annual A100 GPU shipment data is required. However, this information is not publicly disclosed by Nvidia. Consequently, current assessments remain constrained to training-run-level analyses rather than comprehensive industry-scale evaluations.\\
These findings demonstrate that the environmental impact of large-scale AI model training extends beyond operational energy and carbon emissions. The material intensity of hardware production, including the extraction, processing, and disposal of toxic elements, constitutes a significant yet frequently overlooked aspect of AI sustainability \cite{falk2025morethan}. Furthermore, the environmental impact of mining and e-waste disposal is concentrated in regions with limited environmental governance and capacity to mitigate the health and environmental risks associated with exposure to toxic elements \cite{falk2024attribution}. In sub-Saharan Africa, for example, the rapid expansion of mining and processing has increased the concentration of toxic metals in terrestrial, aquatic, and atmospheric systems, thereby exacerbating ecological degradation and health risks for workers and surrounding communities \cite{ripanda202AfricaMining}.

\subsection{Performance Vs Resource Consumption}

\begin{wrapfigure}{r}{0.5\textwidth}
    \centering
\includegraphics[width=1.0\linewidth]{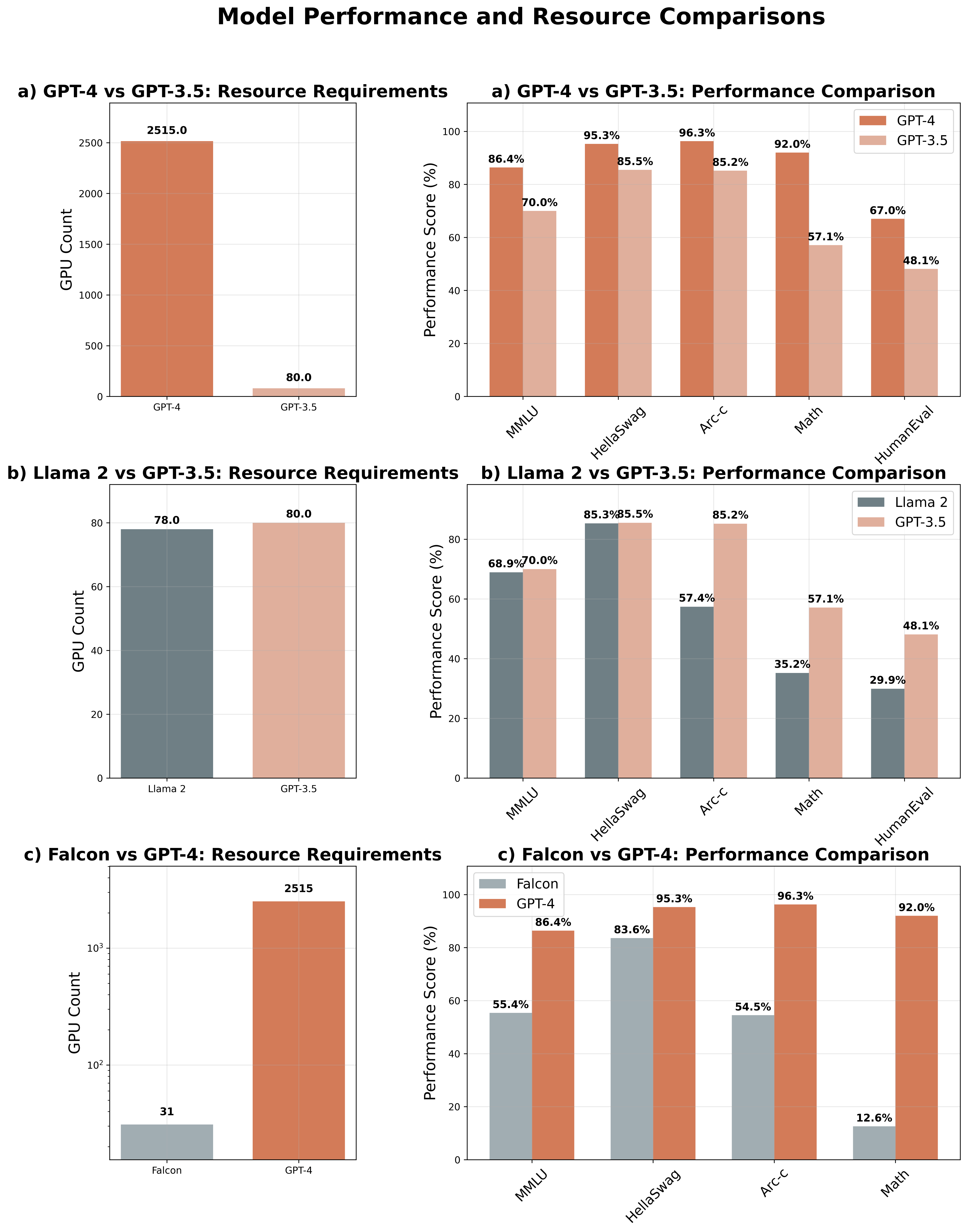}
    \caption{Comparison of model performance between a) GPT-3.5 and GPT-4, b) GPT-3.5 and LLaMa 2, and c) Falcon and GPT-4, across five benchmarks, alongside their respective GPU resource requirement for training (author illustration).}
    \label{fig:performance_comparison}
\end{wrapfigure}

The AI research community has developed a range of standardized benchmarks to systematically monitor technical progress in AI model capabilities over time \cite{stanford2025ai}. Despite inherent limitations, these benchmarks serve as practical tools for evaluating discrete intelligent competencies such as image classification accuracy or multiple-choice question answering. This analysis examines the relationship between benchmark performance and GPU resource requirements, focusing on the following five widely used evaluation frameworks:

\begin{itemize}
    \item  \textit{MATH (mathematical reasoning):} MATH serves as a widely adopted benchmark for evaluating mathematical problem-solving skills based on 12,500 challenging competition mathematics problems. Each problem in MATH has a full step-by-step solution which can be used to teach models to generate answer derivations and explanations \cite{hendrycks2021measuringMathProb}.
    \item  \textit{MMLU (multidisciplinary knowledge):} the Massive Multitask Language Understanding (MMLU) benchmark encompasses 57 tasks, including elementary mathematics, US history, computer science, law, and additional domains \cite{hendrycks2021massiveML}.
    \item \textit{HumanEval (programming proficiency):} HumanEval measures functional correctness in program synthesis from docstrings, comprising 164 original programming problems that assess language comprehension, algorithmic thinking, and mathematical reasoning comparable to entry-level software engineering interviews \cite{chen2021evaluating}. 
    \item \textit{ARC-c (multidisciplinary knowledge):} The AI2's Reasoning Challenge (ARC-c) dataset presents multiple-choice questions derived from science examinations spanning grades 3-9, with the challenge partition containing complex problems requiring advanced reasoning capabilities \cite{clark2018think}.
    \item \textit{HellaSwag (commonsense understanding):} HellaSwag tests a model's commonsense reasoning via natural language inference, focusing on plausible sentence completions in everyday scenarios \cite{zellers2019hellaswag}.  
\end{itemize}
\noindent
\begin{figure} [H]
    \centering
    \includegraphics[width=0.75\linewidth]{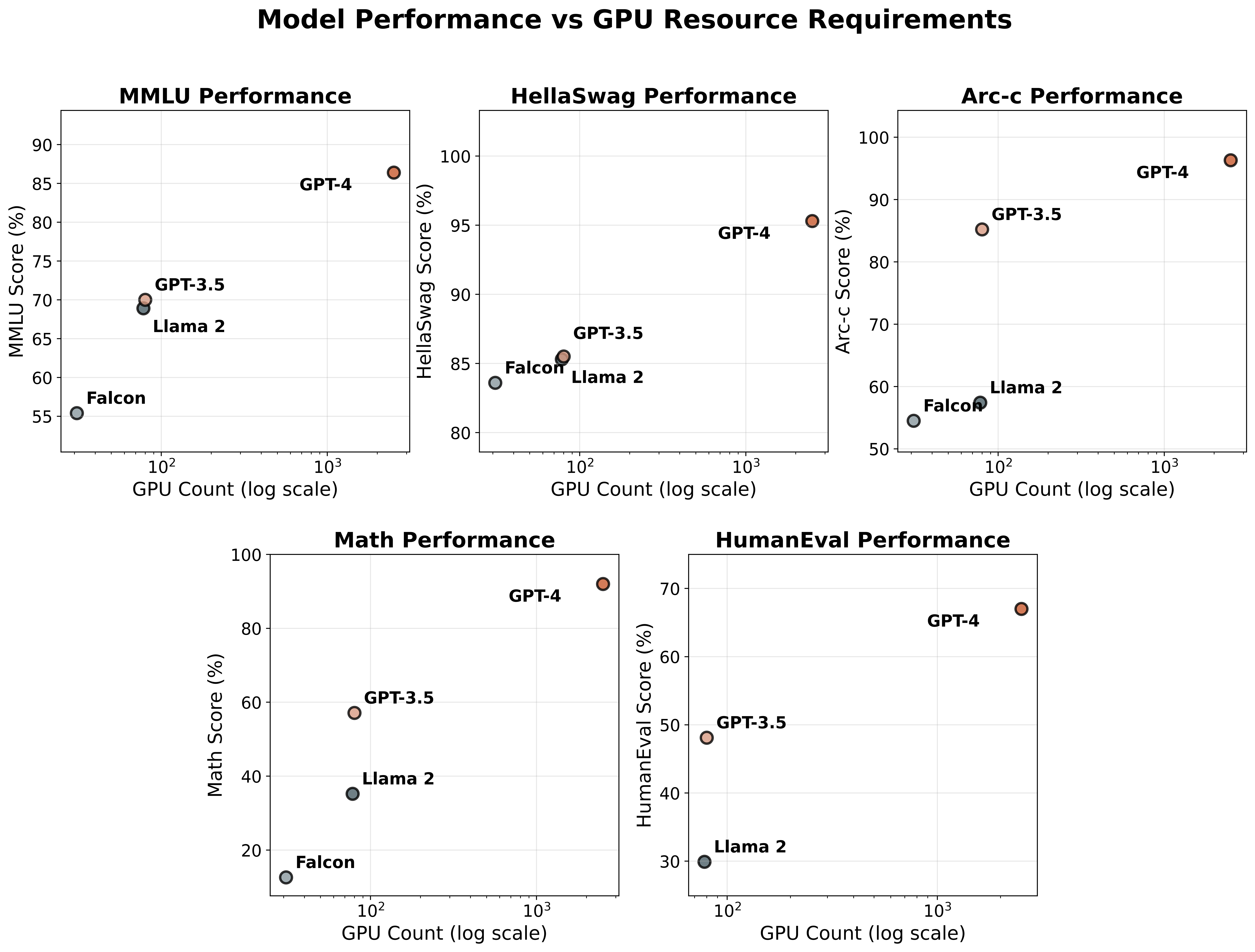}
    \caption{Illustration of the relationship between Falcon, Llama 2, GPT-3.5, and GPT-4 model performance  - measured across five standard benchmarks - and the corresponding GPU requirements (on log scale) for model training (author illustration).}
    \label{fig:scatter_plot}
\end{figure}
\noindent
To better understand the relationship between computational investments and AI performance, we compare models along three dimensions. First, we examine the transition from GPT-3.5 to GPT-4, both developed by OpenAI, to assess the impact of scaling large-scale models within a consistent organizational context. Second, we compare two models trained with nearly identical computational budgets to isolate differences in training efficiency and model performance. Lastly, we contrast the second-smallest with the largest model in our dataset to explore how performance scales at the extremes of model size and resource consumption.\\
To examine in-house performance improvements over time, we first compare OpenAI's GPT-3.5 with its successor, GPT-4 (see Figure \ref{fig:scatter_plot} and \ref{fig:performance_comparison} a)). 
The transition from GPT-3.5 to GPT-4 illustrates substantial computational scaling accompanied by mixed performance returns. GPT-4 required approximately 31.5 times more GPU resources for training than GPT-3.5 (2,515 vs 80 GPUs), representing a more than 3,000\% increase in computational resources. While GPT-4 delivered substantial performance improvements in certain domains, achieving +61.1\% over GPT-3.5 on the MATH benchmark and +39.3\% on HumanEval, other benchmarks showed only modest gains. Overall, these results suggest diminishing returns in terms of performance relative to computational investment. This raises critical questions about the efficiency and sustainability of current scaling trends and whether performance evaluation benchmarks are already saturated.\\
\clearpage

\noindent
The computational demands of training large-scale AI models have increased significantly in recent years. The most straightforward way to measure this is the number of FLOPs required to train an AI model. In 2020, only 11 models required more than 10$^{23}$ FLOPs to train \cite{rahman2025tracking}. However, between January 2020 and June 2025, 432 notable AI models entered the market \cite{epochai2025data}. Among these, 271 models provided estimates of their training FLOPs, while 160 did not. Notably, 111 models among those with available data, roughly 41\%, exceed the 10$^{23}$ FLOPs threshold during training. This trend reflects a significant escalation in the computational scale of modern AI model development.
While increasing computational resources has enabled notable advancement in some domains, it also underscores the limitations of brute-forcing intelligence.\\
To further explore the relationship between compute efficiency and model performance, we compare models trained with similar computational budgets above the 10$^{23}$ FLOPs threshold. This comparison helps isolate factors that contribute to performance increase beyond sheer scaling. Meta's LLaMa 2 (8.4 x 10$^{23}$  Flops) and OpenAI's GPT-3.5  (3.15 x 10$^{23}$ Flops) utilize nearly identical training resources yet demonstrate contrasting efficiency profiles across hardware utilization and model performance evaluations (see Figure \ref{fig:performance_comparison} b)).
LLaMa 2 achieved a higher MFU of 53\% \cite{pytorchHighPerfLlama2}, compared to an estimated 20 to 35\% MFU for GPT-3.5, based on OpenAI's reported 19.6\% MFU for GPT-3 and about 35\% MFU for GPT-4 \cite{daan2024estimating}.
This suggests that Meta's training infrastructure and optimization strategies were more effective in extracting computational throughput from the hardware. However, despite this efficiency advantage and comparable computational budgets, GPT-3.5 consistently outperforms LLaMa 2 across all evaluated benchmarks (see Figure \ref{fig:performance_comparison}), especially in mathematical reasoning and programming tasks. 
This contrast illustrates an important distinction: models trained with similar computational resources can achieve substantially different performance outcomes depending on factors such as training data quality, model architectural design choices, and other optimization strategies (e.g., post-training reinforcement learning). GPT-3.5's stronger performance in mathematical and coding domains, despite lower MFU, may indicate that OpenAI's training data curation and model architecture were better suited for these specific capabilities, even if their training process was less computationally efficient.\\
The comparison between the second smallest model in this study, Falcon (2.4 x 10$^{23}$ Flops), and the largest model, GPT-4 (1.73 x 10$^{25}$ FLOPs), illustrates the relationship between massive resource investment and capability increase (see Figure \ref{fig:performance_comparison} c)). GPT-4 required approximately 81 times the computational resources of Falcon during training. Again, mathematical reasoning capabilities showed the most dramatic improvement, with GPT-4 achieving more than 7 times the performance of Falcon on the math benchmark. This substantial gain suggests that mathematical reasoning represents a particularly resource-intensive capability that may justify extreme computational investments for specific applications. Conversely, commonsense understanding capabilities, as measured by HellaSwag, demonstrate only a modest improvement of 14\%, suggesting that distinct cognitive abilities exhibit differential scaling efficiency in response to increased computational resources.

\subsection{Resource Savings via Training Efficiency and Lifespan Improvements}

\begin{figure} 
    \centering
    \includegraphics[width=0.5\linewidth]{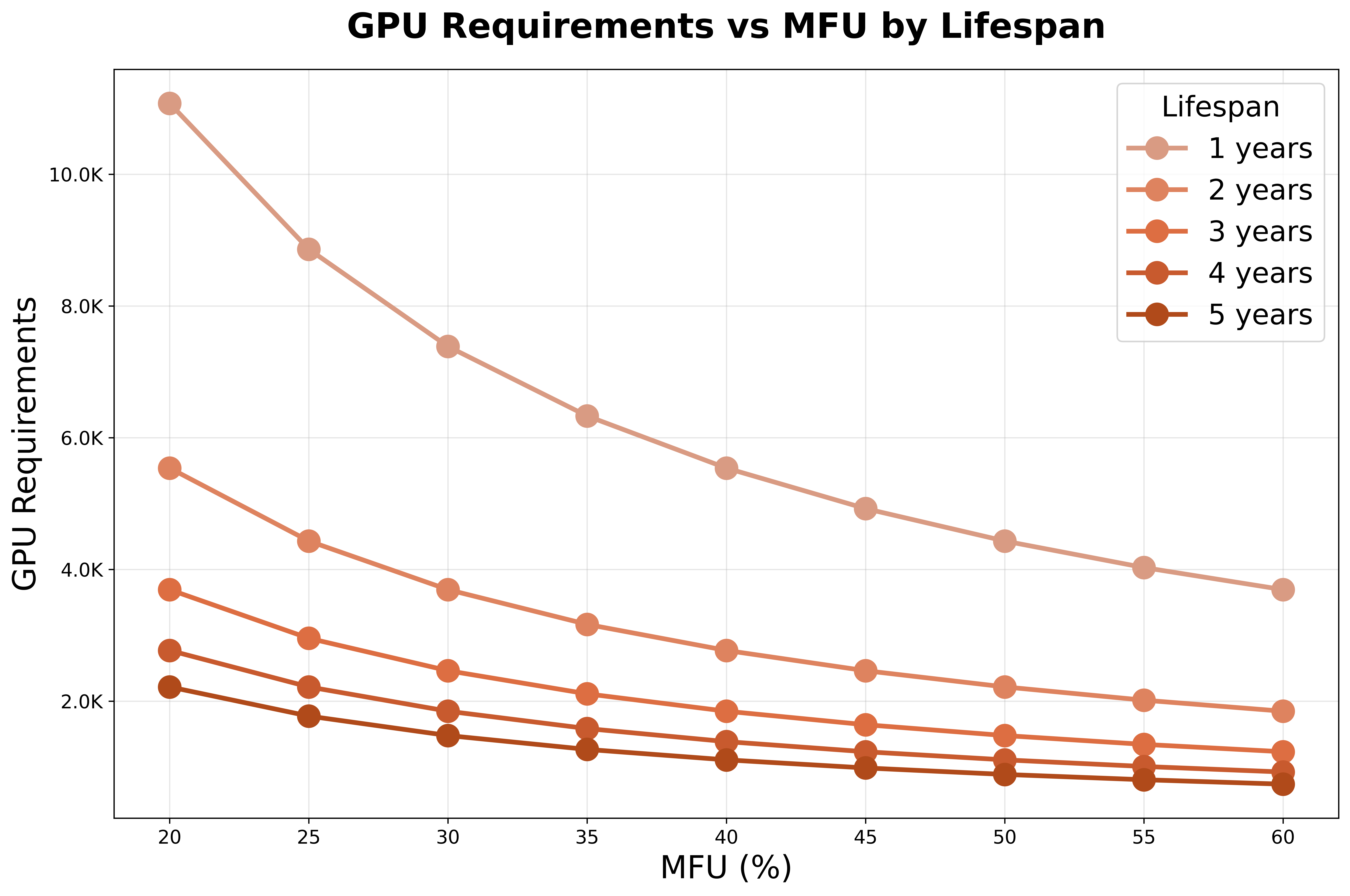}
    \caption{GPU requirements for training GPT-4 for varying MFU scenarios and varying life spans (author illustration).}
    \label{fig:MFUvsLifespan}
\end{figure}

ML engineers, data center operators, and semiconductor manufacturers all play a crucial role in reducing the material footprint of AI training. Two primary strategies can reduce the number of GPUs required for model training: software-based improvements, such as those that optimize GPU utilization; and hardware-based measures, maximizing the lifespan of GPUs in data centers. Data center design, particularly cooling efficiency, has a substantial influence on GPU wear and longevity. These considerations begin at the semiconductor level, where chip designers work to improve thermal management. Prolonged high-energy workloads, common during AI training, can all lead to significant heat generation on silicon chips, accelerating hardware degradation, and ultimately catastrophic failure \cite{tang2024brief, Ycui_2021_flexible}.\\
Increasing the MFU from 20\% to 60\%, while keeping the GPU lifespan constant, reduces the number of GPUs required for model training by approximately 67\%. Similarly, a lifespan expansion from 1 to 3 years, while keeping MFU constant, results in about a 67\% reduction. A further lifespan extension to 5 years would yield an estimated 80\% reduction. To illustrate the impact of these combined optimizations: training GPT-4 with a relatively low MFU of 20\% over a one-year lifespan requires 8,800 GPUs. In contrast, under an optimized scenario, a five-year lifespan and a 60\% MFU would require only 587 GPUs. This represents a potential reduction of about 93\% in GPU usage (see Figure \ref{fig:MFUvsLifespan}). Notably, the decline in GPU requirements with increasing MFU is steeper for shorter lifespans and becomes more gradual for longer lifespans. This trend indicates diminishing returns in GPU savings from MFU improvements once hardware longevity is maximized (see Figure \ref{fig:MFUvsLifespan}).

\section{Discussion}

By examining the performance and resource consumption of AI models, our findings reveal three critical insights for AI development strategies. First, the field appears to be approaching efficiency barriers where pure computational scaling yields diminishing returns for most capabilities. Second, architectural innovations and training methodologies may offer more effective performance improvements than simply scaling raw resources; for example, data quality can play a bigger role than simply scaling the number of tokens. Third, specific capabilities, such as mathematical reasoning, may require disproportional computational investments. This finding aligns with the statement from the benchmark developers in mathematical reasoning, who note that although scaling Transformer models enhances performance on most text-based tasks, it does not currently systematically yield significant improvements in AI models' mathematical capabilities \cite{hendrycks2021measuringMathProb}. To put it differently, we are trying to make a `square' perform the function of a `circle' by applying sheer force and scale.\\
While this study establishes a novel quantitative link between AI's computational demands and its material footprint, some methodological limitations should be considered when interpreting the results. We model GPU demand using a sequential training scenario to standardize comparisons across models with heterogeneous training strategies. This approach is not intended to mirror real-world training workflows, which rely on large-scale parallelization. For example, GPT-4 was reportedly trained on approximately 25,000 A100 GPUs for 90–100 days \cite{katerinaptrvGPT4Leaked}. Under a two-year lifespan assumption, such a training run corresponds to roughly 13\% wear of each GPU's productive capacity. The sequential framing serves as a standardized metric for comparing material intensity across models with different training strategies, translating parallel computational workloads into equivalent hardware utilization. While this framework simplifies system-level complexities, including heterogeneous wear patterns and infrastructure inefficiencies, it captures the fractional depletion of hardware operational lifetime attributable to individual training processes. In addition, MFU alone does not fully characterize training efficiency or hardware wear.
Distributed training introduces communication overheads, heterogeneous parallelization strategies, and deliberate efficiency trade-offs. Lower MFU values may result from strategic training design choices rather than inefficiency. For instance, moderately sized batches distributed across a larger GPU cluster can shorten overall training time even as per-device utilization decreases.\\
The standard FLOP-estimation formula defined in Equation \ref{eq:compute_budget} provides a first-order approximation but omits critical implementation details that affect real-world performance. Performance can vary due to optimized kernel implementations \cite{dao2023flashattention, hsu2025ligerkernel}, re-materialization techniques \cite{chen2016training}, and architectural variations such as sparse models \cite{fedus2021switch}, which can either improve the throughput or reduce the required compute. These factors can introduce deviations from the theoretical FLOP count. While Equation \ref{eq:compute_budget} has its limitations, these estimates remain helpful for comparing the relative computational complexity of different models \cite{hoffmann2022training} and can serve as a conservative baseline for resource allocation and computational needs. Future work could be complemented with empirical measurements that account for hardware utilization patterns, parallelization strategies, and implementation-specific optimizations from real training runs.
\\
As GPT-4's architecture is not officially disclosed, our analysis adopted an informed active parameter estimate based on the most widely cited independent assessments. Alternative configurations, such as different expert counts or activation ratios, would yield different computational and material requirements. To accommodate this uncertainty, we provide an interactive tool\footnote{The tool can be accessed via the following URL: \href{https://huggingface.co/spaces/sophia-falk/flops-2-footprints}{huggingface.co/spaces/sophia-falk/flops-2-footprints}.} that enables users to input alternative parameters and compute the corresponding results themselves.\\
In addition, analytical constraints affect the characterization of the material. The ICP-OES analysis detected 32 elements, but excluded carbon lost during the polymer's pyrolysis and oxygen bound in metal oxides. Unmeasured elements account for 1.78\% of the GPU die, 1.73\% of the heatsink, 7.55\% of the PoP, and 4.05\% of the PCB. Furthermore, the analysis captures only the materials in the final GPU assembly, not losses incurred during material refining, processing, or component manufacturing. This leads to conservative estimates of total resource requirements. Moreover, our assessment focuses exclusively on the GPU unit. Broader data center infrastructure, such as networking equipment, server racks, storage arrays, cooling systems, and power supply, is excluded. Including these infrastructure components would substantially increase AI's estimated material footprint.
Recognizing these limitations is essential for refining future assessments of AI's material consumption and for developing system-level methodologies that more comprehensively capture the resource intensity of computational infrastructures.\\
Ultimately, the material sustainability of AI training depends on a multi-stakeholder approach spanning the entire technology stack. While our analysis focuses on hardware utilization metrics and component lifespan as levers for resource reduction, advancing toward more sustainable AI will require coordinated action across multiple domains, including greater transparency from data center operators and detailed utilization disclosures from AI developers to integrate real-world measurements into future assessments. 
By quantifying the material footprint of AI, this study expands the ongoing sustainability debate beyond energy and water consumption to the physical resource dependencies of computation.
These findings underscore the need to implement both software- and hardware-level optimization strategies to mitigate the environmental and material impacts of large-scale AI training.

\section{Conclusion}

This study provides an integrated assessment linking the computational requirements of modern large-scale AI training to its underlying material footprint. By quantifying the resources required to train contemporary frontier AI models, we demonstrate that the environmental impacts of AI extend beyond operational energy use and further encompass substantial mineral extraction and disposal of toxic materials embedded in advanced computing hardware. Our results further indicate that the performance gains from continued computational scaling are diminishing across a wide range of capability domains. Hence, architectural and algorithmic innovations may offer more sustainable pathways for advancing model performance than brute-forcing intelligence through sheer model scaling. While methodological uncertainties remain, this work establishes a conservative baseline for understanding the physical resource dependencies of AI. 
Reducing these impacts will require coordinated action across developers, hardware manufacturers, and data-center operators, including greater transparency around training practices, standardized reporting of hardware utilization, and better integration of real-world performance data into environmental assessments.
Overall, this study highlights the need to incorporate material considerations into the discussions of AI scalability and sustainability. As AI systems continue to expand in scale and adoption, integrating material impact assessments is crucial for guiding more environmental sustainable development of AI.

\section*{Acknowledgment}
The authors would like to thank ADEME for their material and financial support as well as the cloud providers who supplied broken GPUs for this study. We also extend thanks to Julien Comel, Head of Research at Terra Nova Développement, who conducted the elemental composition analysis.\\
Funded by the TRA Sustainable Futures (University of Bonn) as part of the Excellence Strategy of the federal and state governments.\\
Funding for this research was provided by the Alexander von Humboldt Foundation in the framework of the Alexander von Humboldt Professorship for Artificial Intelligence and endowed by the Federal Ministry of Research to Prof. Dr. Aimee van Wynsberghe.\\

\pagebreak

\printbibliography

@misc{nasir2025world,
  title={The world is running out of resources for IT},
  author={Z. Nasir},
  howpublished = {\url{https://circularcomputing.com/news/the-world-is-running-out-of-resources-for-it/}},
  year={},
  publisher={Circular Computing},
  addendum = {Accessed on 10 April 2025}
}

@misc{sweney2021global,
  title={Global shortage in computer chips 'reaches crisis point'},
  author={M. Sweney},
  howpublished = {\url{https://www.theguardian.com/business/2021/mar/21/global-shortage-in-computer-chips-reaches-crisis-point}},
  year={2021},
  publisher={The Guardian},
  addendum = {Accessed on 10 April 2025}
}

@misc{kharpal2024surging,
  title={Surging AI demand could cause the world's next chip shortage, research says},
  author={A. Kharpal},
  howpublished = {\url{https://www.cnbc.com/2024/09/25/surging-ai-demand-could-cause-the-worlds-next-chip-shortage-report.html}},
  year={2024},
  publisher={CNBC},
  addendum = {Accessed on 10 April 2025}
}

@techreport{unctad2020digital,
  title={Digital economy growth and mineral resources: implications for developing countries},
  author={United Nations Conference on Trade and Development},
  institution={United Nations Conference on Trade and Development},
  type={Technical Note},
  number={TN/UNCTAD/ICT4D/16},
  series={UNCTAD Technical Notes on ICT for Development},
  volume={16},
  pages={44},
  year={2020}
}

@misc{unep2019gobbling,
  title={We're gobbling up the Earth's resources at an unsustainable rate},
  author={UNEP},
  howpublished = {\url{https://www.unep.org/news-and-stories/story/were-gobbling-earths-resources-unsustainable-rate}},
  year={2019},
  publisher={UNEP},
  addendum = {Accessed on 09 April 2025}
}

@misc{unep2024bend,
  title={Bend the trend. Pathways to a liveable planet as resource use spikes},
  author={United Nations Environment Programme},
  howpublished = {\url{https://wedocs.unep.org/20.500.11822/44901}},
  year={2024},
  publisher={United Nations Environment Programme},
  addendum = {Accessed on 10 April 2025}
}

@misc{iea2025executive,
  title={Executive summary – The Role of Critical Minerals in Clean Energy Transitions – Analysis},
  author={IEA},
  howpublished = {\url{https://www.iea.org/reports/the-role-of-critical-minerals-in-clean-energy-transitions/executive-summary}},
  year={},
  publisher={IEA},
  addendum = {Accessed on 13 May 2025}
}

@misc{uddin2025big,
  title={Big Tech lines up over \$300bn in AI spending for 2025},
  author={R. Uddin and S. Morris},
  howpublished = {\url{https://www.ft.com/content/634b7ec5-10c3-44d3-ae49-2a5b9ad566fa}},
  year={2025},
  publisher={Financial Times},
  addendum = {Accessed on 11 April 2025}
}

@misc{mckinsey2025ai,
  title={AI power: Expanding data center capacity to meet growing demand},
  author={McKinsey \& Company},
  howpublished = {\url{https://www.mckinsey.com/industries/technology-media-and-telecommunications/our-insights/ai-power-expanding-data-center-capacity-to-meet-growing-demand}},
  year={2024},
  publisher={McKinsey \& Company},
  addendum = {Accessed on 24 March 2025}
}

@misc{iea2024energy,
  title={Energy and AI. World Energy Outlook Special Report},
  howpublished = {\url{https://iea.blob.core.windows.net/assets/86ed1178-4d77-45ac-ab38-28e849f3b93f/EnergyandAI.pdf}},
  year={2025},
  addendum = {Accessed on 11 April 2025}
}

@misc{spencer2025what,
  title={What the data centre and AI boom could mean for the energy sector},
  author={T. Spencer and S. Singh},
  howpublished = {\url{https://www.iea.org/commentaries/what-the-data-centre-and-ai-boom-could-mean-for-the-energy-sector}},
  year={2024},
  publisher={IEA},
  addendum = {Accessed on 24 March 2025}
}

@misc{iea2024world,
  title={World Energy Outlook 2024},
  author={IEA},
  howpublished = {\url{https://www.iea.org/reports/world-energy-outlook-2024}},
  year={2024},
  publisher={IEA},
  address={Paris},
  addendum = {Accessed on 10 April 2025}
}

@misc{medium2025small,
  title={Small Language Models (SLMs): The Future of Efficient AI},
  author={Medium},
  howpublished = {\url{https://medium.com/@ianishgoswami/small-language-models-slms-the-future-of-efficient-ai-653b033ab7e5}},
  year={2024},
  publisher={Medium},
  addendum = {Accessed on 24 March 2025}
}

@misc{nvidia2020ampere,
  title={nvidia-ampere-architecture-whitepaper.pdf},
  howpublished = {\url{https://images.nvidia.com/aem-dam/en-zz/Solutions/data-center/nvidia-ampere-architecture-whitepaper.pdf}},
  year={2020},
  addendum = {Accessed on 24 March 2025}
}

@misc{leswing2023meet,
  title={Meet the \$10,000 Nvidia chip powering the race for A.I.},
  author={K. Leswing},
  howpublished = {\url{https://www.cnbc.com/2023/02/23/nvidias-a100-is-the-10000-chip-powering-the-race-for-ai-.html}},
  year={2023},
  publisher={CNBC},
  addendum = {Accessed on 24 March 2025}
}

@misc{heath2024mark,
  title={Mark Zuckerberg's new goal is creating artificial general intelligence},
  author={A. Heath},
  howpublished = {\url{https://www.theverge.com/2024/1/18/24042354/mark-zuckerberg-meta-agi-reorg-interview}},
  year={2024},
  publisher={The Verge},
  addendum = {Accessed on 24 March 2025}
}

@misc{eu2024regulation,
  title={Regulation (EU) 2024/1689 of the European Parliament and of the Council of 13 June 2024 laying down harmonised rules on artificial intelligence and amending Regulations (EC) No 300/2008, (EU) No 167/2013, (EU) No 168/2013, (EU) 2018/858, (EU) 2018/1139 and (EU) 2019/2144 and Directives 2014/90/EU, (EU) 2016/797 and (EU) 2020/1828 (Artificial Intelligence Act) (Text with EEA relevance)},
  howpublished = {\url{http://data.europa.eu/eli/reg/2024/1689/oj/eng}},
  year={2024},
  publisher={The European Parliament},
  addendum = {Accessed on 24 March 2025}
}

@article{grattafiori2024llama,
  title={The Llama 3 Herd of Models},
  author={A. Grattafiori and others},
  journal={arXiv preprint arXiv:2407.21783},
  year={2024},
  doi={10.48550/arXiv.2407.21783}
}

@misc{ohiri2025nvidia,
  title={NVIDIA A100 vs. V100: In-Depth GPU Comparison},
  author={E. Ohiri},
  howpublished = {\url{https://www.cudocompute.com/blog/nvidia-a100-vs-v100-how-do-they-compare}},
  year={2024},
  publisher={CUDO Compute},
  addendum = {Accessed on 21 March 2025}
}

@misc{cybersided2025how,
  title={How Long Do GPUs Last? (Average Lifespan \& Effectiveness)},
  author={Cybersided},
  howpublished = {\url{https://cybersided.com/how-long-do-gpus-last/}},
  year={},
  publisher={Cybersided},
  addendum = {Accessed on 21 March 2025}
}

@inproceedings{ostrouchov2020gpu,
  title={GPU Lifetimes on Titan Supercomputer: Survival Analysis and Reliability},
  author={G. Ostrouchov and D. Maxwell and R. A. Ashraf and C. Engelmann and M. Shankar and J. H. Rogers},
  booktitle={SC20: International Conference for High Performance Computing, Networking, Storage and Analysis},
  pages={1--14},
  year={2020},
  doi={10.1109/SC41405.2020.00045}
}

@misc{techfund2024ai,
  title={AI architect at Google mentions the lifetime of datacenter GPUs at current utilization levels to be 3 years at most. This is v bullish for \$NVDA end-demand as most analysts were assuming a lifetime of around 5 years},
  author={Tech Fund [@techfund1]},
  howpublished = {\url{https://x.com/techfund1/status/1849031571421983140}},
  year={2024},
  publisher={Twitter},
  addendum = {Accessed on 21 March 2025}
}

@misc{shilov2025datacenter,
  title={Datacenter GPU service life can be surprisingly short — only one to three years is expected according to unnamed Google architect},
  author={A. Shilov},
  howpublished = {\url{https://www.tomshardware.com/pc-components/gpus/datacenter-gpu-service-life-can-be-surprisingly-short-only-one-to-three-years-is-expected-according-to-unnamed-google-architect}},
  year={2024},
  publisher={Tom's Hardware},
  addendum = {Accessed on 21 March 2025}
}

@misc{bahdanau2025flops,
  title={The FLOPs Calculus of Language Model Training},
  author={D. Bahdanau},
  howpublished = {\url{https://medium.com/@dzmitrybahdanau/the-flops-calculus-of-language-model-training-3b19c1f025e4}},
  year={2022},
  publisher={Medium},
  addendum = {Accessed on 24 March 2025}
}

@misc{rahman2025tracking,
  title={Tracking Large-Scale AI Models},
  author={R. Rahman, David Owen, Josh You},
  howpublished = {\url{https://epoch.ai/blog/tracking-large-scale-ai-models}},
  year={2024},
  publisher={Epoch AI},
  addendum = {Accessed on 24 March 2025}
}

@misc{epochai2025data,
  title={Data on Notable AI Models},
  author={Epoch AI},
  howpublished = {\url{https://epoch.ai/data/notable-ai-models}},
  year={2025},
  publisher={Epoch AI},
  addendum = {Accessed on 05 June 2025}
}

@article{kaplan2020scaling,
  title={Scaling Laws for Neural Language Models},
  author={J. Kaplan and others},
  journal={arXiv preprint arXiv:2001.08361},
  year={2020},
  doi={10.48550/arXiv.2001.08361}
}

@article{hoffmann2022training,
  title={Training Compute-Optimal Large Language Models},
  author={J. Hoffmann and others},
  journal={arXiv preprint arXiv:2203.15556},
  year={2022},
  doi={10.48550/arXiv.2203.15556}
}

@misc{katerinaptrvGPT4Leaked,
  title={GPT4- All Details Leaked},
  author={katerinaptrv},
  howpublished = {\url{https://medium.com/@daniellefranca96/gpt4-all-details-leaked-48fa20f9a4a}},
  year={2023},
  publisher={Medium},
  addendum = {Accessed on 05 June 2025}
}

@article{chowdhery2022palm,
  title={PaLM: Scaling Language Modeling with Pathways},
  author={A. Chowdhery and others},
  journal={arXiv preprint arXiv:2204.02311},
  year={2022},
  doi={10.48550/arXiv.2204.02311}
}

@misc{pytorchLargeScale,
  title={Large Scale Training of Hugging Face Transformers on TPUs With PyTorch/XLA FSDP},
  howpublished = {\url{https://pytorch.org/blog/large-scale-training-hugging-face/}},
  year={2023},
  publisher={PyTorch},
  addendum = {Accessed on 11 April 2025}
}

@misc{fryepaid,
  title={'I paid for the whole GPU, I am going to use the whole GPU': A high-level guide to GPU utilization},
  author={C. Frye},
  howpublished = {\url{https://modal.com/blog/gpu-utilization-guide?utm}},
  year={2025},
  publisher={Modal},
  addendum = {Accessed on 11 April 2025}
}

@misc{pytorchHighPerfLlama2,
  title={High-Performance Llama 2 Training and Inference with PyTorch/XLA on Cloud TPUs – PyTorch},
  howpublished = {\url{https://pytorch.org/blog/high-performance-llama-2/}},
  year={2023},
  publisher={PyTorch},
  addendum = {Accessed on 23 June 2025}
}

@misc{daan2024estimating,
  title={Estimating efficiency improvements in LLM pre-training},
  author={Daan},
  howpublished = {\url{https://www.lesswrong.com/posts/tJAD2LG9uweeEfjwq/estimating-efficiency-improvements-in-llm-pre-training}},
  year={2024},
  addendum = {Accessed on 11 April 2025}
}

@misc{pytorchMaximizing,
  title={Maximizing training throughput using PyTorch FSDP – PyTorch},
  howpublished = {\url{https://pytorch.org/blog/maximizing-training/}},
  year={2024},
  publisher={PyTorch},
  addendum = {Accessed on 04 June 2025}
}

@misc{behindthekeyboardLlama,
  title={llama/MODEL\_CARD.md at main · BehindTheKeyboard/llama},
  author={BehindTheKeyboard},
  howpublished = {\url{https://github.com/BehindTheKeyboard/llama/blob/main/MODEL_CARD.md}},
  year={2024},
  publisher={GitHub},
  addendum = {Accessed on 24 June 2025}
}

@article{tang2024brief,
  title={Brief overview of the impact of thermal stress on the reliability of through silicon via: Analysis, characterization, and enhancement},
  author={S. Tang and others},
  journal={Mater. Sci. Semicond. Process.},
  volume={183},
  pages={108745},
  year={2024},
  doi={10.1016/j.mssp.2024.108745}
}

@article{Ycui_2021_flexible,
  title={Flexible thermal interface based on self-assembled boron arsenide for high-performance thermal management},
  author={Y. Cui and Z. Qin and H. Wu and M. Li and Y. Hu},
  journal={Nature Communication},
  volume={12},
  number={1},
  pages={1284},
  year={2021},
  doi={10.1038/s41467-021-21531-7}
}

@misc{nvidia2017volta,
  title={volta-architecture-whitepaper.pdf},
  howpublished = {\url{https://images.nvidia.com/content/volta-architecture/pdf/volta-architecture-whitepaper.pdf}},
  year={2017},
  addendum = {Accessed on 24 June 2025}
}

@article{han2023big,
  title={The Big Chip: Challenge, model and architecture},
  author={Y. Han and others},
  journal={Fundam. Res.},
  volume={4},
  number={6},
  pages={1431--1441},
  year={2023},
  doi={10.1016/j.fmre.2023.10.020}
}

@article{wang2024analysis,
  title={Analysis of soil fertility and toxic metal characteristics in open-pit mining areas in northern Shaanxi},
  author={N. Wang and Z. Liu and Y. Sun and N. Lu and Y. Luo},
  journal={Sci. Rep.},
  volume={14},
  number={1},
  pages={2273},
  year={2024},
  doi={10.1038/s41598-024-52886-8}
}

@article{ripanda202AfricaMining,
  title={Combatting toxic chemical elements pollution for Sub-Saharan Africa's ecological health},
  author={Ripanda, Asha and Hossein, Miraji and Rwiza, Mwemezi J and Nyanza, Elias Charles and Selemani, Juma Rajabu and Nkrumah, Salma and Bakari, Ramadhani and Alfred, Mateso Said and Machunda, Revocatus L and Vuai, Said Ali Hamad},
  journal={Environmental Pollution and Management},
  year={2025},
  publisher={Elsevier}
}

@article{dutta2021IndustrialWastewater,
  title={Industrial wastewater treatment: Current trends, bottlenecks, and best practices},
  author={Dutta, Deblina and Arya, Shashi and Kumar, Sunil},
  journal={Chemosphere},
  year={2021},
  publisher={Elsevier}
}

@article{wu2022antimonyBeCO,
  title={Antimony, beryllium, cobalt, and vanadium in urban park soils in Beijing: Machine learning-based source identification and health risk-based soil environmental criteria},
  author={Wu, Yihang and Liu, Qiyuan and Ma, Jin and Zhao, Wenhao and Chen, Haiyan and Qu, Yajing},
  journal={Environmental Pollution},
  volume={293},
  pages={118554},
  year={2022},
  publisher={Elsevier}
}

@book{WHO2010toxic,
  title={World health statistics 2010},
  author={World Health Organization},
  year={2010},
  publisher={World Health Organization}
}

@article{falk2024attribution,
  title={The attribution problem of a seemingly intangible industry},
  author={S. Falk and A. Van Wynsberghe and L. Biber-Freudenberger},
  journal={Environmental Challenges},
  volume={16},
  pages={101003},
  year={2024},
  doi={10.1016/j.envc.2024.101003}
}

@misc{stanford2025ai,
  title={The 2025 AI Index Report | Stanford HAI},
  howpublished = {\url{https://hai.stanford.edu/ai-index/2025-ai-index-report}},
  year={2025},
  addendum = {Accessed on 23 June 2025}
}

@article{hendrycks2021measuringMathProb,
  title={Measuring Mathematical Problem Solving With the MATH Dataset},
  author={Hendrycks, Dan and Burns, Collin and Kadavath, Saurav and Arora, Akul and Basart, Steven and Tang, Eric and Song, Dawn and Steinhardt, Jacob},
  journal={arXiv preprint arXiv:2103.03874},
  year={2021},
  doi={10.48550/arXiv.2103.03874}
}

@article{hendrycks2021massiveML,
  title={Measuring Massive Multitask Language Understanding},
  author={Hendrycks, Dan and Burns, Collin and Basart, Steven and Zou, Andy and Mazeika, Mantas and Song, Dawn and Steinhardt, Jacob},
  journal={arXiv preprint arXiv:2009.03300},
  year={2021},
  doi={10.48550/arXiv.2009.03300}
}

@article{chen2021evaluating,
  title={Evaluating Large Language Models Trained on Code},
  author={Chen, Mark and Tworek, Jerry and Jun, Heewoo and Yuan, Qiming and Pinto, Henrique Ponde De Oliveira and Kaplan, Jared and Edwards, Harri and Burda, Yuri and Joseph, Nicholas and Brockman, Greg and others},
  journal={arXiv preprint arXiv:2107.03374},
  year={2021},
  doi={10.48550/arXiv.2107.03374}
}

@article{clark2018think,
  title={Think you have Solved Question Answering? Try ARC, the AI2 Reasoning Challenge},
  author={Clark, Peter and Cowhey, Isaac and Etzioni, Oren and Khot, Tushar and Sabharwal, Ashish and Schoenick, Carissa and Tafjord, Oyvind},
  journal={arXiv preprint arXiv:1803.05457},
  year={2018},
  doi={10.48550/arXiv.1803.05457}
}

@article{zellers2019hellaswag,
  title={HellaSwag: Can a Machine Really Finish Your Sentence?},
  author={R. Zellers and A. Holtzman and Y. Bisk and A. Farhadi and Y. Choi},
  journal={arXiv preprint arXiv:1905.07830},
  year={2019},
  doi={10.48550/arXiv.1905.07830}
}

@article{falk2025morethan,
  title={More than Carbon: Cradle-to-Grave environmental impacts of GenAI training on the Nvidia A100 GPU},
  author={Falk, Sophia and Ekchajzer, David and Pirson, Thibault and Lees-Perasso, Etienne and Wattiez, Augustin and Biber-Freudenberger, Lisa and Luccioni, Sasha and van Wynsberghe, Aimee},
  journal={arXiv preprint arXiv:2509.00093},
  year={2025}
}

@article{falk2025DC_wasteheat,
  title={The Potential of Data Center Waste Heat Recovery for Greenhouse Food Production in the US: Ramifications for Sustainable AI},
  author={Falk, Sophia and Asgari, Nima and Pearce, Joshua M and van Wynsberghe, Aimee},
  journal={Available at SSRN 5170348}
}

@inproceedings{lang2024comprehensive,
  title={A comprehensive study on quantization techniques for large language models},
  author={Lang, Jiedong and Guo, Zhehao and Huang, Shuyu},
  booktitle={2024 4th International Conference on Artificial Intelligence, Robotics, and Communication (ICAIRC)},
  pages={224--231},
  year={2024},
  organization={IEEE}
}

@inproceedings{
hsu2025ligerkernel,
title={Liger-Kernel: Efficient Triton Kernels for {LLM} Training},
author={Pin-Lun Hsu and Yun Dai and Vignesh Kothapalli and Qingquan Song and Shao Tang and Siyu Zhu and Steven Shimizu and Shivam Sahni and Haowen Ning and Yanning Chen and Zhipeng Wang},
booktitle={Championing Open-source DEvelopment in ML Workshop @ ICML25},
year={2025},
url={https://openreview.net/forum?id=36SjAIT42G}
}

@article{dao2023flashattention,
  title={Flashattention-2: Faster attention with better parallelism and work partitioning},
  author={Dao, Tri},
  journal={arXiv preprint arXiv:2307.08691},
  year={2023}
}

@article{chen2016training,
  title={Training deep nets with sublinear memory cost},
  author={Chen, Tianqi and Xu, Bing and Zhang, Chiyuan and Guestrin, Carlos},
  journal={arXiv preprint arXiv:1604.06174},
  year={2016}
}

@article{fedus2021switch,
  title={Switch Transformers: Scaling to trillion parameter models with simple and efficient sparsity.(2021)},
  author={Fedus, William and Zoph, Barret and Shazeer, Noam},
  journal={arXiv preprint cs.LG/2101.03961},
  year={2021}
}

@inproceedings{jiang2024megascale,
  title={$\{$MegaScale$\}$: Scaling large language model training to more than 10,000 $\{$GPUs$\}$},
  author={Jiang, Ziheng and Lin, Haibin and Zhong, Yinmin and Huang, Qi and Chen, Yangrui and Zhang, Zhi and Peng, Yanghua and Li, Xiang and Xie, Cong and Nong, Shibiao and others},
  booktitle={21st USENIX Symposium on Networked Systems Design and Implementation (NSDI 24)},
  pages={745--760},
  year={2024}
}

@article{zhao2024efficiently,
  title={Efficiently training 7b llm with 1 million sequence length on 8 gpus},
  author={Zhao, Pinxue and Zhang, Hailin and Fu, Fangcheng and Nie, Xiaonan and Liu, Qibin and Yang, Fang and Peng, Yuanbo and Jiao, Dian and Li, Shuaipeng and Xue, Jinbao and others},
  journal={arXiv e-prints},
  pages={arXiv--2407},
  year={2024}
}

@article{jamil2025emlio,
  title={EMLIO: Minimizing I/O Latency and Energy Consumption for Large-Scale AI Training},
  author={Jamil, Hasibul and Nine, MD and Kosar, Tevfik},
  journal={arXiv preprint arXiv:2508.11035},
  year={2025}
}

@article{go2025characterizing,
  title={Characterizing the Efficiency of Distributed Training: A Power, Performance, and Thermal Perspective},
  author={Go, Seokjin and Park, Joongun and More, Spandan and Wu, Hanjiang and Wang, Irene and Jezghani, Aaron and Krishna, Tushar and Mahajan, Divya},
  journal={arXiv preprint arXiv:2509.10371},
  year={2025}
}

@inproceedings{luccioni2025REBOUND,
  title={From efficiency gains to rebound effects: The problem of jevons' paradox in AI's polarized environmental debate},
  author={Luccioni, Alexandra Sasha and Strubell, Emma and Crawford, Kate},
  booktitle={Proceedings of the 2025 ACM Conference on Fairness, Accountability, and Transparency},
  pages={76--88},
  year={2025}
}

@incollection{jevons2018coal,
  title={The coal question: An Inquiry concerning the Progress of the Nation, and the Probable Exhaustion of our Coal-mines},
  author={Jevons, W Stanley},
  booktitle={The Economics of Population},
  pages={193--204},
  year={1866},
  publisher={Routledge}
}

@misc{openai2024gpt4technicalreport,
      title={GPT-4 Technical Report}, 
      author={OpenAI and Josh Achiam and Steven Adler and Sandhini Agarwal and Lama Ahmad and Ilge Akkaya and Florencia Leoni Aleman and Diogo Almeida and Janko Altenschmidt and Sam Altman and Shyamal Anadkat and Red Avila and Igor Babuschkin and Suchir Balaji and Valerie Balcom and Paul Baltescu and Haiming Bao and Mohammad Bavarian and Jeff Belgum and Irwan Bello and Jake Berdine and Gabriel Bernadett-Shapiro and Christopher Berner and Lenny Bogdonoff and Oleg Boiko and Madelaine Boyd and Anna-Luisa Brakman and Greg Brockman and Tim Brooks and Miles Brundage and Kevin Button and Trevor Cai and Rosie Campbell and Andrew Cann and Brittany Carey and Chelsea Carlson and Rory Carmichael and Brooke Chan and Che Chang and Fotis Chantzis and Derek Chen and Sully Chen and Ruby Chen and Jason Chen and Mark Chen and Ben Chess and Chester Cho and Casey Chu and Hyung Won Chung and Dave Cummings and Jeremiah Currier and Yunxing Dai and Cory Decareaux and Thomas Degry and Noah Deutsch and Damien Deville and Arka Dhar and David Dohan and Steve Dowling and Sheila Dunning and Adrien Ecoffet and Atty Eleti and Tyna Eloundou and David Farhi and Liam Fedus and Niko Felix and Simón Posada Fishman and Juston Forte and Isabella Fulford and Leo Gao and Elie Georges and Christian Gibson and Vik Goel and Tarun Gogineni and Gabriel Goh and Rapha Gontijo-Lopes and Jonathan Gordon and Morgan Grafstein and Scott Gray and Ryan Greene and Joshua Gross and Shixiang Shane Gu and Yufei Guo and Chris Hallacy and Jesse Han and Jeff Harris and Yuchen He and Mike Heaton and Johannes Heidecke and Chris Hesse and Alan Hickey and Wade Hickey and Peter Hoeschele and Brandon Houghton and Kenny Hsu and Shengli Hu and Xin Hu and Joost Huizinga and Shantanu Jain and Shawn Jain and Joanne Jang and Angela Jiang and Roger Jiang and Haozhun Jin and Denny Jin and Shino Jomoto and Billie Jonn and Heewoo Jun and Tomer Kaftan and Łukasz Kaiser and Ali Kamali and Ingmar Kanitscheider and Nitish Shirish Keskar and Tabarak Khan and Logan Kilpatrick and Jong Wook Kim and Christina Kim and Yongjik Kim and Jan Hendrik Kirchner and Jamie Kiros and Matt Knight and Daniel Kokotajlo and Łukasz Kondraciuk and Andrew Kondrich and Aris Konstantinidis and Kyle Kosic and Gretchen Krueger and Vishal Kuo and Michael Lampe and Ikai Lan and Teddy Lee and Jan Leike and Jade Leung and Daniel Levy and Chak Ming Li and Rachel Lim and Molly Lin and Stephanie Lin and Mateusz Litwin and Theresa Lopez and Ryan Lowe and Patricia Lue and Anna Makanju and Kim Malfacini and Sam Manning and Todor Markov and Yaniv Markovski and Bianca Martin and Katie Mayer and Andrew Mayne and Bob McGrew and Scott Mayer McKinney and Christine McLeavey and Paul McMillan and Jake McNeil and David Medina and Aalok Mehta and Jacob Menick and Luke Metz and Andrey Mishchenko and Pamela Mishkin and Vinnie Monaco and Evan Morikawa and Daniel Mossing and Tong Mu and Mira Murati and Oleg Murk and David Mély and Ashvin Nair and Reiichiro Nakano and Rajeev Nayak and Arvind Neelakantan and Richard Ngo and Hyeonwoo Noh and Long Ouyang and Cullen O'Keefe and Jakub Pachocki and Alex Paino and Joe Palermo and Ashley Pantuliano and Giambattista Parascandolo and Joel Parish and Emy Parparita and Alex Passos and Mikhail Pavlov and Andrew Peng and Adam Perelman and Filipe de Avila Belbute Peres and Michael Petrov and Henrique Ponde de Oliveira Pinto and Michael and Pokorny and Michelle Pokrass and Vitchyr H. Pong and Tolly Powell and Alethea Power and Boris Power and Elizabeth Proehl and Raul Puri and Alec Radford and Jack Rae and Aditya Ramesh and Cameron Raymond and Francis Real and Kendra Rimbach and Carl Ross and Bob Rotsted and Henri Roussez and Nick Ryder and Mario Saltarelli and Ted Sanders and Shibani Santurkar and Girish Sastry and Heather Schmidt and David Schnurr and John Schulman and Daniel Selsam and Kyla Sheppard and Toki Sherbakov and Jessica Shieh and Sarah Shoker and Pranav Shyam and Szymon Sidor and Eric Sigler and Maddie Simens and Jordan Sitkin and Katarina Slama and Ian Sohl and Benjamin Sokolowsky and Yang Song and Natalie Staudacher and Felipe Petroski Such and Natalie Summers and Ilya Sutskever and Jie Tang and Nikolas Tezak and Madeleine B. Thompson and Phil Tillet and Amin Tootoonchian and Elizabeth Tseng and Preston Tuggle and Nick Turley and Jerry Tworek and Juan Felipe Cerón Uribe and Andrea Vallone and Arun Vijayvergiya and Chelsea Voss and Carroll Wainwright and Justin Jay Wang and Alvin Wang and Ben Wang and Jonathan Ward and Jason Wei and CJ Weinmann and Akila Welihinda and Peter Welinder and Jiayi Weng and Lilian Weng and Matt Wiethoff and Dave Willner and Clemens Winter and Samuel Wolrich and Hannah Wong and Lauren Workman and Sherwin Wu and Jeff Wu and Michael Wu and Kai Xiao and Tao Xu and Sarah Yoo and Kevin Yu and Qiming Yuan and Wojciech Zaremba and Rowan Zellers and Chong Zhang and Marvin Zhang and Shengjia Zhao and Tianhao Zheng and Juntang Zhuang and William Zhuk and Barret Zoph},
      year={2024},
      eprint={2303.08774},
      archivePrefix={arXiv},
      primaryClass={cs.CL},
      url={https://arxiv.org/abs/2303.08774}, 
}

@article{patel2023ScalingWall,
  title={The ai brick wall--a practical limit for scaling dense transformer models, and how gpt 4 will break past it},
  author={Patel, Dylan},
  journal={SemiAnalysis. January},
  volume={24},
  pages={2023},
  year={2023}
}

@article{patelWong2023GPT-4Architecture,
  title={GPT-4 Architecture, Infrastructure, Training Dataset, Costs, Vision, MoE
},
  author={Patel, Dylan and Wong, Gerald},
  journal={SemiAnalysis. July},
  volume={},
  pages={},
  howpublished = {\url{https://newsletter.semianalysis.com/p/gpt-4-architecture-infrastructure}},
  year={2023},
  addendum = {Accessed on 11 Nov 2025}
}

@article{Erdil2024MoEinference,
  title={How do mixture-of-experts models compare to dense models in inference?},
  author={Erdil, Ege},
  journal={Epoch.ai},
  volume={},
  pages={},
  howpublished = {\url{https://epoch.ai/gradient-updates/moe-vs-dense-models-inference}},
  year={2024},
  addendum = {Accessed on 11 Nov 2025}
}

@article{Fischer2025MoEExplained,
  title={LLM Mixture of Experts Explained},
  author={Fischer Lemos, Claudio},
  journal={tensorops.ai},
  volume={},
  pages={},
  howpublished = {\url{https://www.tensorops.ai/post/what-is-mixture-of-experts-llm}},
  year={2025},
  addendum = {Accessed on 11 Nov 2025}
}

@article{chen2025AIenergy,
  title={How much energy will AI really consume? The good, the bad and the unknown},
  author={Chen, Sophia},
  journal={Nature},
  volume={639},
  number={8053},
  pages={22--24},
  year={2025},
  publisher={Nature}
}

@article{deVries2023growing,
  title={The growing energy footprint of artificial intelligence},
  author={De Vries, Alex},
  journal={Joule},
  volume={7},
  number={10},
  pages={2191--2194},
  year={2023},
  publisher={Elsevier}
}

@inproceedings{luccioni2024powerHungry,
  title={Power hungry processing: Watts driving the cost of AI deployment?},
  author={Luccioni, Sasha and Jernite, Yacine and Strubell, Emma},
  booktitle={Proceedings of the 2024 ACM conference on fairness, accountability, and transparency},
  pages={85--99},
  year={2024}
}

@article{li2025makingAIlessthirty,
  title={Making ai less' thirsty'},
  author={Li, Pengfei and Yang, Jianyi and Islam, Mohammad A and Ren, Shaolei},
  journal={Communications of the ACM},
  volume={68},
  number={7},
  pages={54--61},
  year={2025},
  publisher={ACM New York, NY, USA}
}

@article{zhang2021DCcooling,
  title={A survey on data center cooling systems: Technology, power consumption modeling and control strategy optimization},
  author={Zhang, Qingxia and Meng, Zihao and Hong, Xianwen and Zhan, Yuhao and Liu, Jia and Dong, Jiabao and Bai, Tian and Niu, Junyu and Deen, M Jamal},
  journal={Journal of Systems Architecture},
  volume={119},
  pages={102253},
  year={2021},
  publisher={Elsevier}
}

@misc{SemiAnalysis2023GPT-4Archi,
  title={GPT-4 Architecture, Infrastructure, Training Dataset, Costs, Vision, MoE},
  author={Dylan Patel and Gerald Wong},
  howpublished = {\url{https://newsletter.semianalysis.com/p/gpt-4-architecture-infrastructure}},
  year={2023},
  publisher={SemiAnalysis},
  addendum = {Accessed on 19 Nov 2025}
}

@article{Xiao2025aiNetZeroPath,
  title={Environmental impact and net-zero pathways for sustainable artificial intelligence servers in the USA},
  author={Xiao, Tianqi and Fuso Nerini, Francesco and Matthews, H. Damon and Tavoni, Massimo and You, Fengqi},
  journal={Nature Sustainability},
  doi={https://doi.org/10.1038/s41893-025-01681-y},
  year={2025},
  pages={1--13},
  publisher={Nature}
}

\end{document}